\documentclass[12pt, letterpaper]{article}
\usepackage[margin=2.5cm]{geometry}
\usepackage{amsmath}
\usepackage{epsfig}
\usepackage[cp1252]{inputenc}
\usepackage{amsfonts}
\usepackage{amssymb}
\usepackage{setspace}
\usepackage{authblk}
\usepackage{epsfig}
\usepackage{xcolor}
\usepackage{makeidx}
\usepackage{subfig}
\usepackage{cite}
\usepackage{titling}
\usepackage{lastpage}
\usepackage[nolists, nomarkers]{endfloat}
\usepackage{fancyhdr}
\usepackage{amsthm}

\usepackage{titlesec}
\usepackage[font={small,sl}]{caption}

\titleformat{\section}
{\normalfont\bfseries\scshape}{\thesection}{1em}{}

\titleformat{\subsection}
{\normalfont\bfseries\scshape}{\thesubsection}{1em}{}

\titleformat{\subsubsection}
{\normalfont\slshape}{\thesubsubsection}{1em}{}

\titleformat{\title}{\normalfont\bfseries}{\thesection}{1em}{}

\usepackage{hyperref}
\usepackage[normalem]{ulem}
\newcommand{\sV}{\ensuremath{V_0}}
\newcommand{\waveP}{\ensuremath{c_+}}
\newcommand{\waveM}{\ensuremath{c_{-}}}

\newcommand{\markthis}[1]{\textcolor{red}{\textbf{#1}}}

\title{\large \bfseries Opposite moving detachment waves mediate stick-slip friction at soft interfaces\vspace{-2em}}
\author{\normalsize Mohammad Aaquib Ansari}
\author{\normalsize Koushik Viswanathan\thanks{Email: koushik@iisc.ac.in}\vspace{-1em}}
\affil{\normalsize \textsl{Dept. of Mechanical Engineering, Indian Institute of Science, Bangalore}}

\date{\normalsize \textsc{\today}\vspace{-3em}}
\begin{document}
\maketitle
\hrulefill
\begin{abstract}
  Intermittent motion, called stick--slip, is a friction instability that commonly occurs during relative sliding of two elastic solids. In adhesive polymer contacts, where elasticity and interface adhesion are strongly coupled, stick--slip results from the propagation of slow detachment waves at the interface. Using \emph{in situ} imaging experiments at an adhesive contact, we show the occurrence of two distinct detachment waves moving parallel (Schallamach wave) and anti-parallel (separation wave) to the applied remote sliding. Both waves cause slip in the same direction and travel at speeds much lesser than any elastic wave speed. We use an elastodynamic framework to describe the propagation of these slow detachment waves at an elastic-rigid interface and obtain governing integral equations in the low wave speed limit. These integral equations are solved in closed form when the elastic solid is incompressible. Two solution branches emerge, corresponding to opposite moving detachment waves, just as seen in the experiments. A numerical scheme is used to obtain interface stresses and velocities for the incompressible case for arbitrary Poisson ratio. Based on these results, we explicitly demonstrate a correspondence between propagating slow detachment waves and a static bi-material interface crack. Based on this, and coupled with a recently proposed fracture analogy for dynamic friction, we develop a phase diagram showing domains of possible occurrence of stick--slip via detachment waves vis-\'a-vis steady interface sliding. 
\end{abstract}

\section{Introduction}

Consider a simple system consisting of two rectangular solid blocks, one elastic the other relatively rigid, that are pressed into contact and slid remotely at constant velocity $V_0$. Elementary considerations dictate that the interface will start sliding once the shear stress exceeds the static friction threshold. This threshold is commonly assumed to depend on the coefficient of friction $\mu_s$ and the normal force $F_N$. That this rudimentary picture is simply not universally true is borne out dramatically by earthquake faults, squealing brakes and violin strings \cite{BowdenLeben_ProcRoySocA_1939, Rabinowicz_SciAm_1956, BraceByerlee_Science_1966, BowdenTabor_Friction_1973}. These systems exhibit what is known as stick--slip, a phenomenon wherein the interface moves only intermittently even though the contacting solids are remotely moved at constant relative speed. Several plausible explanations for stick--slip have now been established, all of them hinging primarily on velocity-dependence of the interface friction force \cite{Rabinowicz_SciAm_1956, Dieterich_JGeophysRes_1979, Ruina_JGeophysRes_1983, Scholz_Nature_1998} and/or some type of regularization \cite{CochardRice_JGeophysRes_2000, RanjithRice_JMechPhysSolids_2001}.

However, soft material contacts can exhibit stick--slip via more subtle mechanisms \cite{BaumbergerCaroli_AdvPhys_2006}. This complication is primarily because friction, adhesion and elastic deformation cannot be decoupled. Under many conditions, soft adhesive interfaces can slip only via the propagation of pulse-like rupture fronts, often involving local interface detachment and demonstrating unique dynamics \cite{BaumbergerETAL_PhysRevLett_2002, YamaguchiETAL_JGeoPhysRes_2011}. A well-known example of such a detachment wave is the Schallamach wave in rubbers \cite{Schallamach_Wear_1971}, which results from local interface buckling \cite{Barquins_Wear_1993, KoudineBarquins_JAdhSciTech_1996, RandCrosby_ApplPhysLett_2006, FukahoriETAL_Wear_2010, ViswanathanETAL_PhysRevE_2015}. 

Detachment waves of this nature are often described by analogy with the motion of a ruck in a carpet \cite{VellaETAL_PhysRevLett_2009, KolinskiETAL_PhysRevLett_2009}: if a carpet is to be moved by unit distance $\Delta x$, we could either simply translate the entire carpet surface at once by $\Delta x$, or create a localized slip zone by buckling---causing unit slip $\Delta x$---that then propagates along the carpet and progressively causes it to slip. This buckling and detachment type mechanism is quite general and also occurs during motion of soft-bodied insects such as caterpillars and earthworms \cite{GrayLissman_JExptBio_1938, Trueman_SoftBodiedLocomotion_1975}.

In soft polymer interfaces, moving detachment waves are characterized by their dramatically low propagation velocity \cite{Schallamach_Wear_1971, Barquins_Wear_1985, FukahoriETAL_Wear_2010, ViswanathanETAL_SoftMatter_2016_1, BaumbergerETAL_PhysRevLett_2002}. In fact, this characteristic makes it difficult to theoretically describe interface waves in soft polymers. On the contrary, fast moving waves are directly described by linear elastodynamics so that one can compute interface displacements, velocities and stresses \cite{Achenbach_WavePropagation_2012, Adams_JApplMech_1998, ComninouDundurs_JApplMech_1978}. The precise boundary conditions at the interface determine the propagation velocity, but it is always comparable to the Rayleigh wave speed \cite{Stoneley_ProcRoySocA_1924, AchenbachEpstein_JEnggMechDiv_1967, ComninouDundurs_JApplMech_1977}, which is a material property. 

%This is quite unlike analogous interface waves in hard materials, which are known to move at speeds comparable to the shear wave speed of the material \cite{RubinsteinETAL_Nature_2004}x
Consequently, several questions about the propagation of slow detachment waves at soft interfaces remain unanswered and are the subject of the present manuscript. Firstly, what types of detachment waves can occur and what determines the corresponding wave velocity? Secondly, how is unit slip due to the passage of a single wave related to its propagation direction? Finally, motivated by a recently described correspondence between friction and fracture \cite{RubinsteinETAL_Nature_2004, SvetlizkyETAL_PhysRevLett_2017}, can we establish a relationship between detachment wave motion and crack growth in soft interfaces? Such a relationship can then be used to predict when detachment waves, and consequently stick--slip, will occur at the expense of steady uniform sliding. 

To be able to answer these questions, we first experimentally investigate the nature of detachment waves at a model adhesive sliding interface between a soft polymer and a hard glass. Using high-speed \emph{in situ} imaging, we show that two opposite moving detachment waves can propagate at the interface, all the while causing interface slip in the direction of remote $V_0$. We establish the properties of these waves based on multiple experimental observations. A general theory is then presented describing slow-moving elastic waves, starting from linear elastodynamics and ending in a pair of coupled singular integral equations. In the incompressible limit $(\nu =0.5)$, we show that these equations can be solved exactly. Based on physical kinematic constraints, two independent solution branches emerge, describing opposite-moving waves. Numerical solutions for the general case ($\nu \neq 0.5$) are presented, along with corresponding interface velocities and stresses. This reveals some fundamental similarities between detachment wave motion and bi-material interface fracture. Coupled with a fracture analogy for the onset of dynamic friction, we develop a phase diagram that shows domains of possible occurrence of stick--slip via detachment waves vis-\'a-vis steady sliding. 

\section{Experimental configuration}

\begin{figure}
  \centering
  \includegraphics[width=\textwidth]{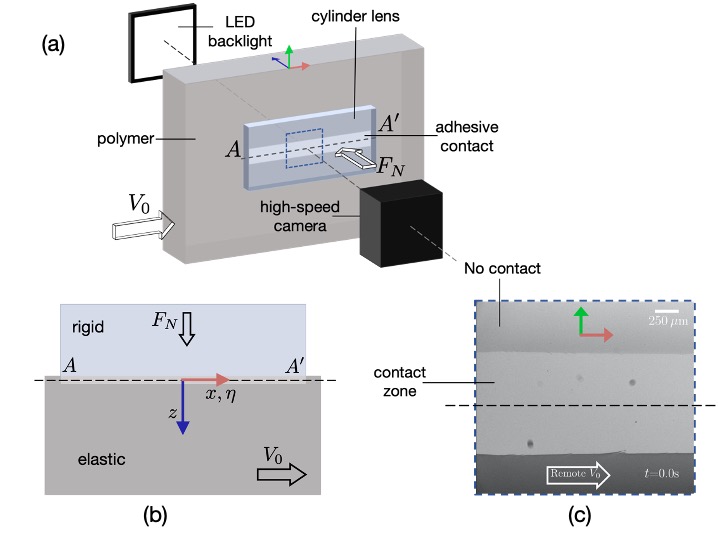}
  \caption{Experimental setup and coordinate conventions used in the text. (a) Schematic showing polymer-lens contact geometry and sliding conditions. The $xyz$ axes are also shown as red, green and blue arrows, respectively. (b) Approximate 2D side view of the interface; the comoving coordinate $\eta$ coincides with $x$ (see text) and (c) sample camera image showing adhesive contact zone (light grey) distinguished from the rest of the polymer (dark grey). Remotely applied $\sV$ is always taken to be from left to right, along the $x$-axis. }
  \label{fig:exptSchematic}
\end{figure}

Quantitative information about the interface dynamics is obtained using a model adhesive interface that is slid at low velocity $\sV < 1$ mm/s, see Fig.~\ref{fig:exptSchematic}. A polydimethylsiloxane (PDMS, Dow Corning Sylgard 184) slab is brought into contact with a cylindrical glass lens (Edmund Optics). The PDMS slab is prepared following standard procedures---by mixing base (vinyl--terminated polydimethylsiloxane) with a curing agent (methylhydrosiloxane--dimethylsiloxane copolymer) in the ratio 10:1 by weight. The resulting mixture is cured for 6 hours at 100$^\circ$C and then at room temperature ($\sim 30^\circ$C) for 18 hours. The mould setup and other preparation procedures are identical to those described in earlier work, see Ref.~\cite{ViswanathanETAL_SoftMatter_2016_1}. The resulting PDMS sample had dimensions of 22 mm $\times$ 70 mm $\times$ 25 mm. Young's modulus and Poisson's ratio for PDMS are around 1 MPa and 0.46 respectively, based on shear and bulk modulus values reported in the literature \cite{Mark_PolymerDataHandbook_2009}.

The lens used as the rigid indenter was plano-convex with face radius $16.25$ mm and length $25$ mm. The contact geometry and coordinate system are shown in Fig.~\ref{fig:exptSchematic}. The lens and PDMS were brought into line contact (along $x$-axis, length $L = 25$ mm) and pressed together with a normal load $F_N$, adjusted so that the contact width (1 mm) was much lesser than $L$. This \lq adhesive channel\rq\ along the $xy$ plane allowed the isolation of single stick--slip events in contrast to conventionally used spherical contacts that are severely constrained by edge effects. The contact interface is backlit by an LED light source (Metaphase Technologies) and recorded using a high--speed imaging system (PCO dimax) with spatial and temporal resolution of $2.8\,\mu$m per pixel and 0.2 ms, respectively.

The PDMS is mounted on a linear motorized translating axis that can move at constant velocities between $10$ $\mu$m/s and $20$ mm/s. Simultaneous with \emph{in situ} imaging, normal and shear forces were measured using a piezoelectric dynamometer (Kistler 9254). In a typical experiment, the lens and PDMS are first brought into contact and maintained for a fixed time $t=60$ s to standardize any possible contact aging effects. A remote $\sV$ is then applied to the PDMS via the motorized axis for a sliding distance of atleast 30 mm. $F_N$ was nominally measured to be $\sim 50$ mN, with shear forces arising from $\sV$ being an order of magnitude larger.

\section{Two distinct opposite moving detachment waves}

At low sliding velocity, the interface shows rich spatio-temporal dynamics, even though the remote \sV\ is constant for a given experiment. The interface remains stationary for long periods, separated by periods of slip. Slip events are solely mediated by the propagation of detachment waves at the interface. Two distinct waves occur, moving parallel and anti-parallel to the remotely applied \sV\ direction. 

\begin{figure}[ht!]
  \centering
  \includegraphics[width=\textwidth]{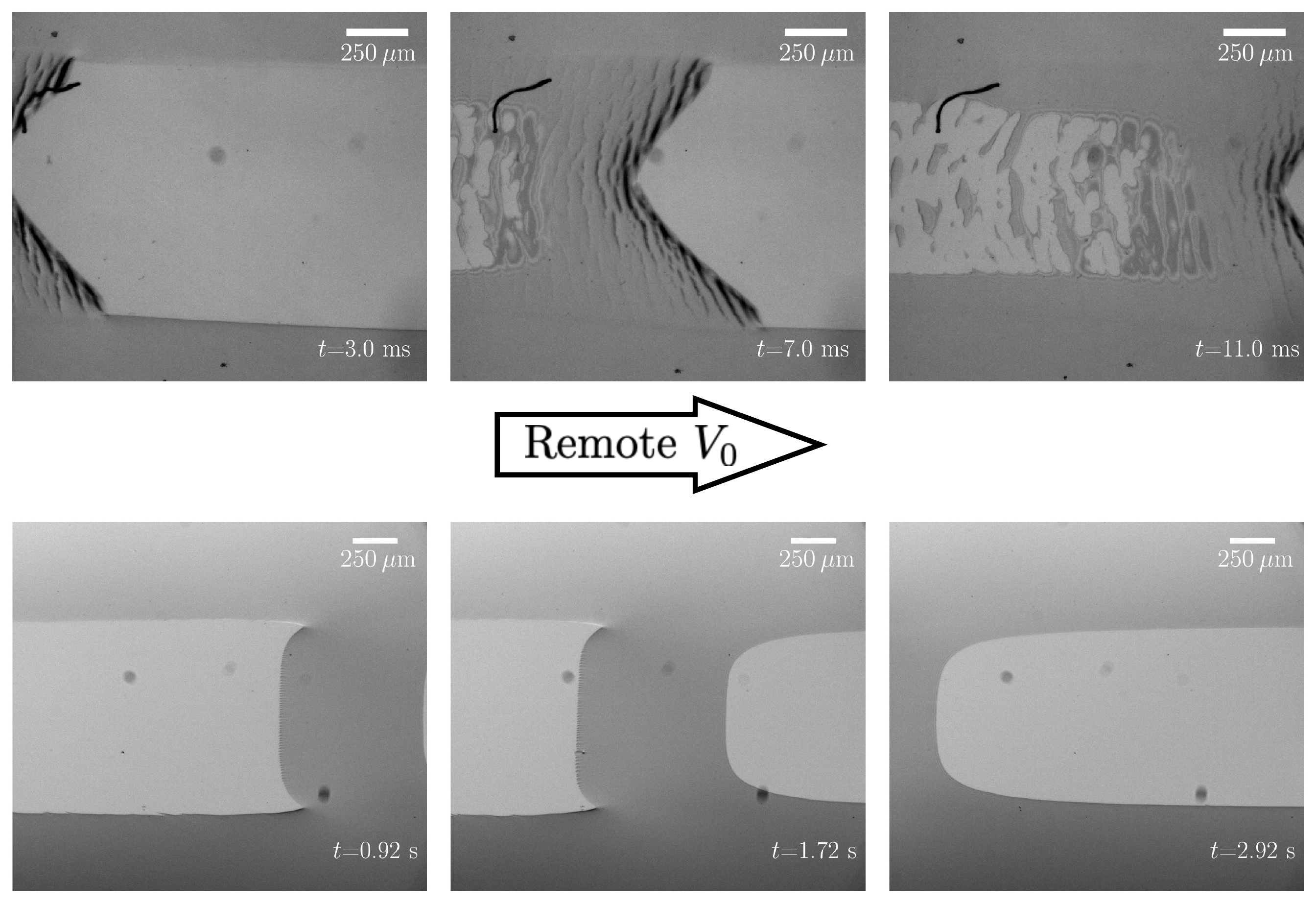}
  \caption{High-speed \emph{in situ} images showing two distinct detachment waves at the polymer-lens interface. Top row shows a Schallamach ($+$) wave propagating in the same direction as remotely applied $\sV$. Note the timescale for wave propagation is significantly lesser than that for uniform sliding. Bottom row shows a separation ($-$) wave propagating in the opposite direction. Applied velocity $\sV = $ 0.5 mm/s (top row) and 0.05 mm/s (bottom row).}
  \label{fig:dWaveFrames}
\end{figure}

The cylindrical contact geometry and the resulting adhesive channel help isolate single wave events without any interfering edge effects. As a result, the propagation of two distinct detachment waves is clearly observed, see Fig.~\ref{fig:dWaveFrames}. The top panel shows the motion of a Schallamach or positive d-wave and the bottom panel shows a separation pulse or a negative d-wave. We use the terms positive ($+$) and negative ($-$) to explicitly denote the direction of motion of these two waves vis-\'a-vis applied \sV . Remote sliding direction \sV\ is to the right in the figure as denoted by large arrows. Movie sequences showing individual waves are provided as supplementary material \markthis{M1}. 

We first discuss the top panel in Fig.~\ref{fig:dWaveFrames}. The sequence of three frames in this panel shows the motion of a single Schallamach ($+$) wave within the interface, $\sV = 0.5$ mm/s. Several features of this wave are immediately obvious from the sequence. Firstly, the detachment zone  (dark) shows a characteristic V-shape with wrinkles on its surface. The wrinkles form locally as the wave progresses, and result in imperfect contact in its wake. Secondly, points on the interface remain stationary before and after wave passage; they only translate in the \sV\ direction when the wave propagates past them. This feature is clear by observing the motion of a marker particle that is embedded on the elastomer surface (black in figure). The marker is translated by the $+$ wave so that it slips a unit distance $\Delta x_+$ after the wave has passed. Thirdly, the timestamps show that the wave propagates at a speed $\waveP$ much larger than \sV . It is further clear that $\waveP$ ($\sim 0.1$ m/s) is orders of magnitude lower than any elastic wave speed in the material, \emph{cf.} Rayleigh wave speed $c_R \sim 10^3$ m/s. Finally, once the wave has passed, the entire sequence repeats with the occurrence of another wave. Such single wave events occur at a constant frequency $n_+$ such that $n_+ \Delta x_+  = \sV$. 

The second panel in Fig.~\ref{fig:dWaveFrames} shows a very different type of wave---the separation $(-)$ wave---that propagates in the opposite direction to \sV\ . The $(-)$ wave distinct from, yet also shares some simlarities with, the $+$ wave described above. Firstly, the detachment zone is quite devoid of any features such as wrinkles and folds, and also has a shape quite distinct from the V-shape of the $+$ wave.  Consequently, complete readhesion is seen in the wake of the wave, see frame 2. Secondly, just as with the $+$ wave, points on the interface remain stationary until wave passage. After the wave has passed, surface points are translated by a unit amount $\Delta x_{-}$ in the direction of \sV . It is noteworthy that surface points slip in the same direction as \sV\ even though the wave moves in the opposite direction. Thirdly, timestamps on the frames show that the wave speed $\waveM$ is larger than $\sV$ but $\waveM \ll c_R$ as before. Once a single wave propagates, the entire process repeats and interface motion occurs in steps just as with $+$ waves. Furthermore, $\waveM < \waveP$ and the frequency $n_{-} < n_{+}$.

%Interface dynamics accompanying wave motion is also quantitatively revealed by a space--time diagram, constructed by stacking a horizontal line in each image of a temporal sequence, see Fig.~\ref{fig:spaceTime} (top panel). Such a diagram consequently shows the spatio-temporal history of points along this line within the interface. Diagrams for both the Schallamach ($+$) and separation ($-$) waves are shown in panels (a) and (b) of Fig.~\ref{fig:spaceTime}, respectively. \markthis{[Some sentences describing the ST diagrams here]}

\begin{figure}[ht!]
  \centering
  \begin{minipage}{0.45\textwidth}
  \includegraphics[width=\textwidth]{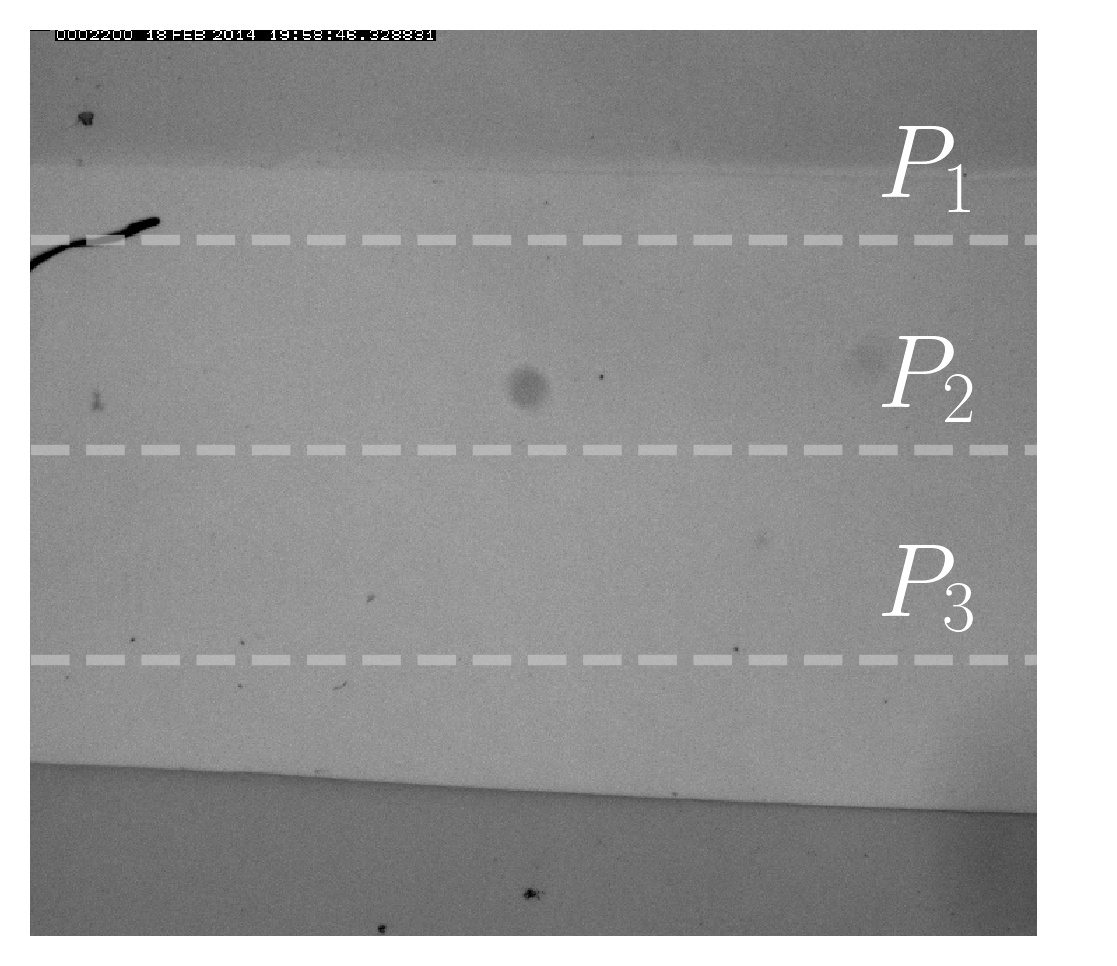}  
  \end{minipage}
  \quad
  \begin{minipage}{0.8\textwidth}
    \includegraphics[width=\textwidth]{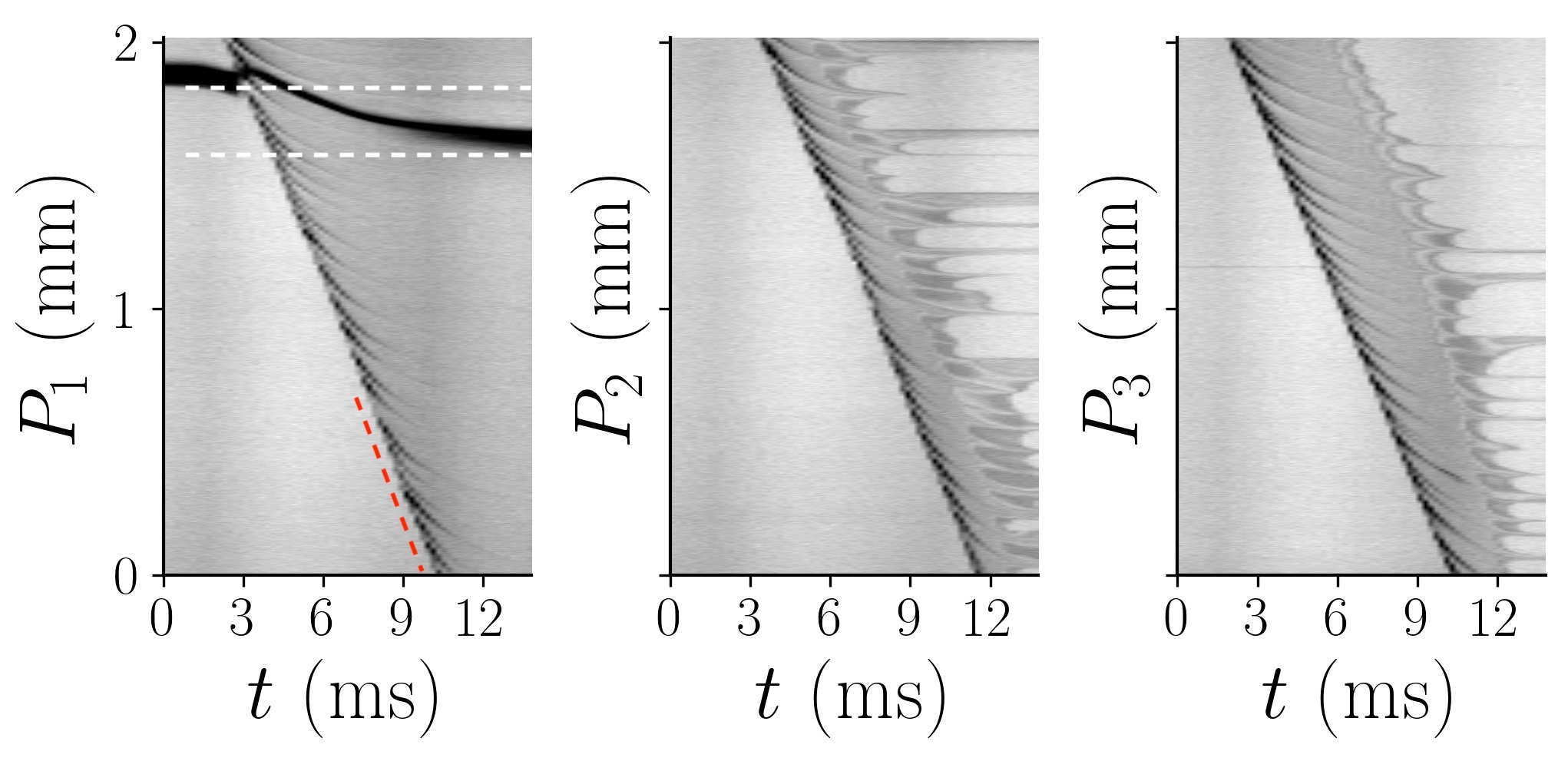}
  \end{minipage}
  \caption{Space--time diagrams showing propagation of Schallamach ($+$) waves. Top panel shows locations of the horizontal lines $P_1, P_2, P_3$ and bottom panel shows the three corresponding space--time diagrams. The wave appears as a dark band in all three diagrams with constant slope (red dashed line) equal to the wave speed. Unit interface slip distance, as indicated y the trajectory of the black line marker, is as shown between parallel white dashed lines in the bottom panel.}%
  \label{fig:spaceTimeScW}
\end{figure}

Interface dynamics accompanying wave motion is quantitively revealed by means of space-time diagrams. These are constructed by stacking a horizontal line in the image sequences as a function of time. Three such stacks, denoted $P_1, P_2, P_3$ for Schallamach ($+$) waves are presented in Fig.~\ref{fig:spaceTimeScW}. The wave itself appears as a dark diagonal band in each space-time diagram, all three bands having the same inclination (red dashed line). Consequently, the $+$ wave has a constant speed as it propagates through the interface. The negative value of the slope is due to the wave moving in the same direction as applied $\sV$. Additionally, the interface slip is also self-evident in frame $P_1$ by following the trace of the dark patch (top panel) in the space-time diagram. The patch has slipped by unit distance (between white dashed lines) due to wave motion.

\begin{figure}[ht!]
  \centering
  \begin{minipage}{0.45\textwidth}
  \includegraphics[width=\textwidth]{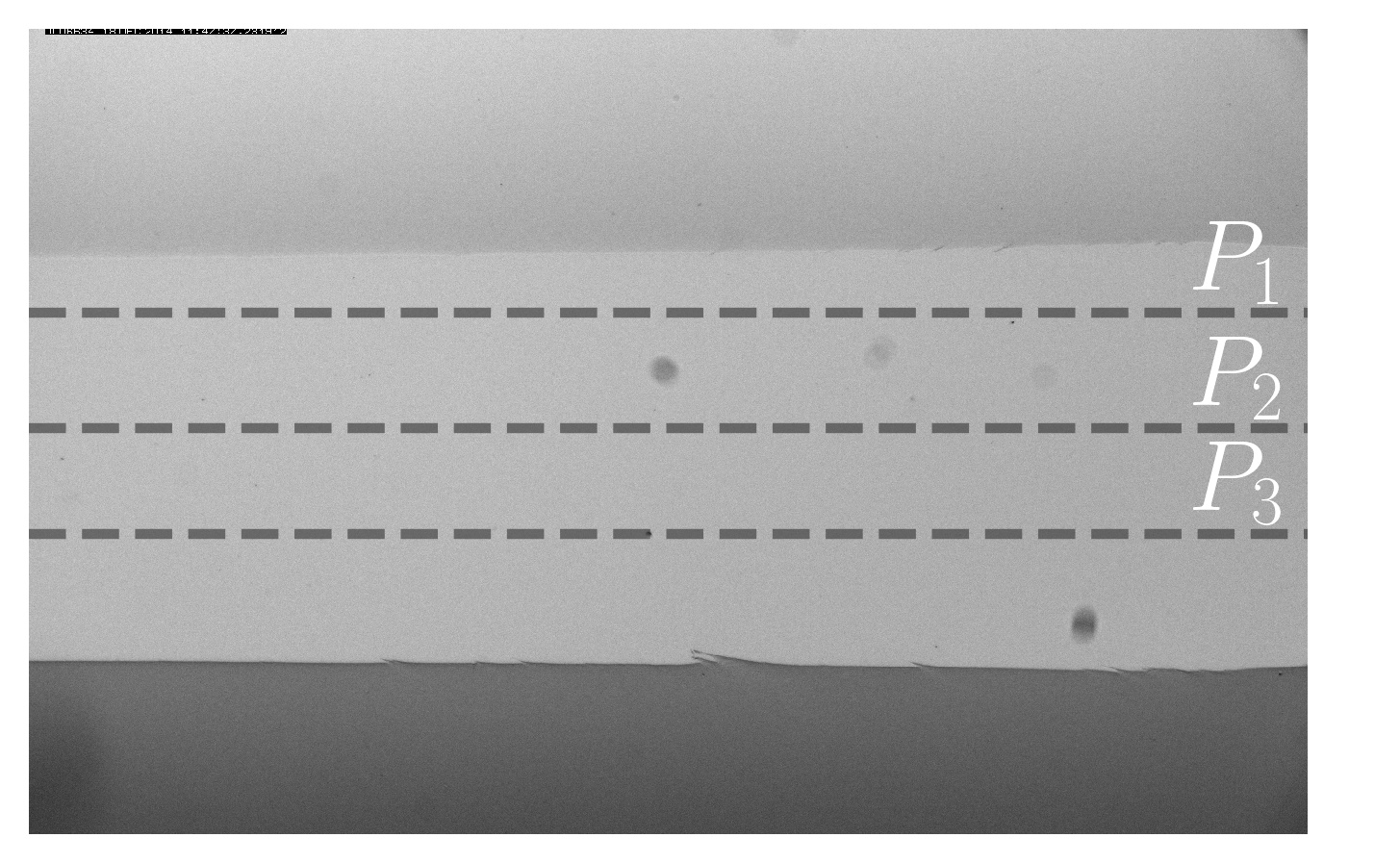}  
  \end{minipage}
  \quad
  \begin{minipage}{0.8\textwidth}
    \includegraphics[width=\textwidth]{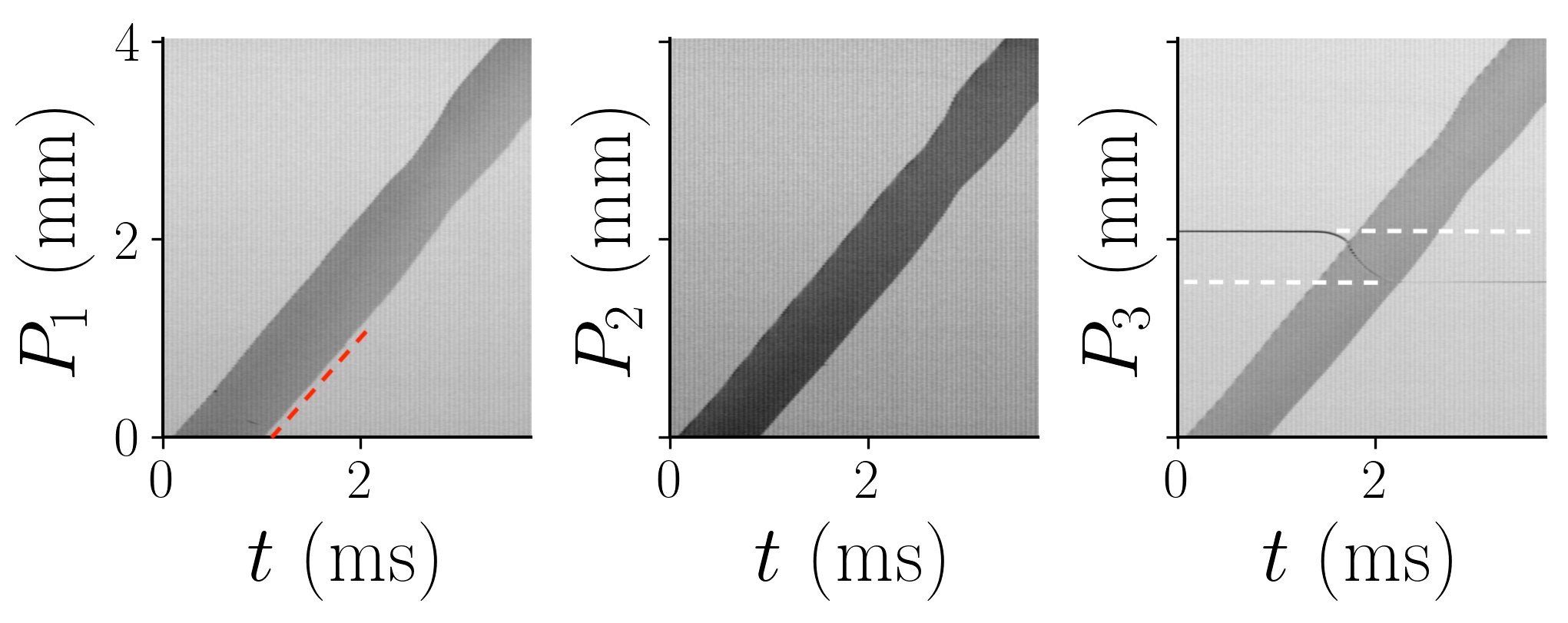}
  \end{minipage}
  \caption{Space--time diagrams showing propagation of separation ($-$) pulses. Top panel shows locations of the horizontal lines $P_1, P_2, P_3$ and bottom panel shows the three corresponding space--time diagrams. The wave appears as a dark band in all three diagrams with constant slope (red dashed line) equal to the wave speed. Unit interface slip distance, as indicated by a moving dirt particle in the bottom panel, is marked by parallel white dashed lines.}%
  \label{fig:spaceTimeSP}
\end{figure}

An analogous space--time diagram can also be constructed for separation ($-$) waves, see Fig.~\ref{fig:spaceTimeSP}. Here again, three horizontal lines $P_1$ to $P_3$ are stacked as a function of time. Just as before, the wave appears as an inclined dark band with constant slope (red dashed line) in all three panels. However, in contrast to the $+$ wave, the band now has a positive slope because it moves in the opposite direction. As with the dark patch in Fig.~\ref{fig:spaceTimeScW}, the time trajectory of a dirt particle on the surface reveals the amount of interface slip due to a single wave (between white dashed lines, last panel).  

\begin{figure}[ht!]
  \centering
  \begin{minipage}{0.48\textwidth}
  \includegraphics[width=\textwidth]{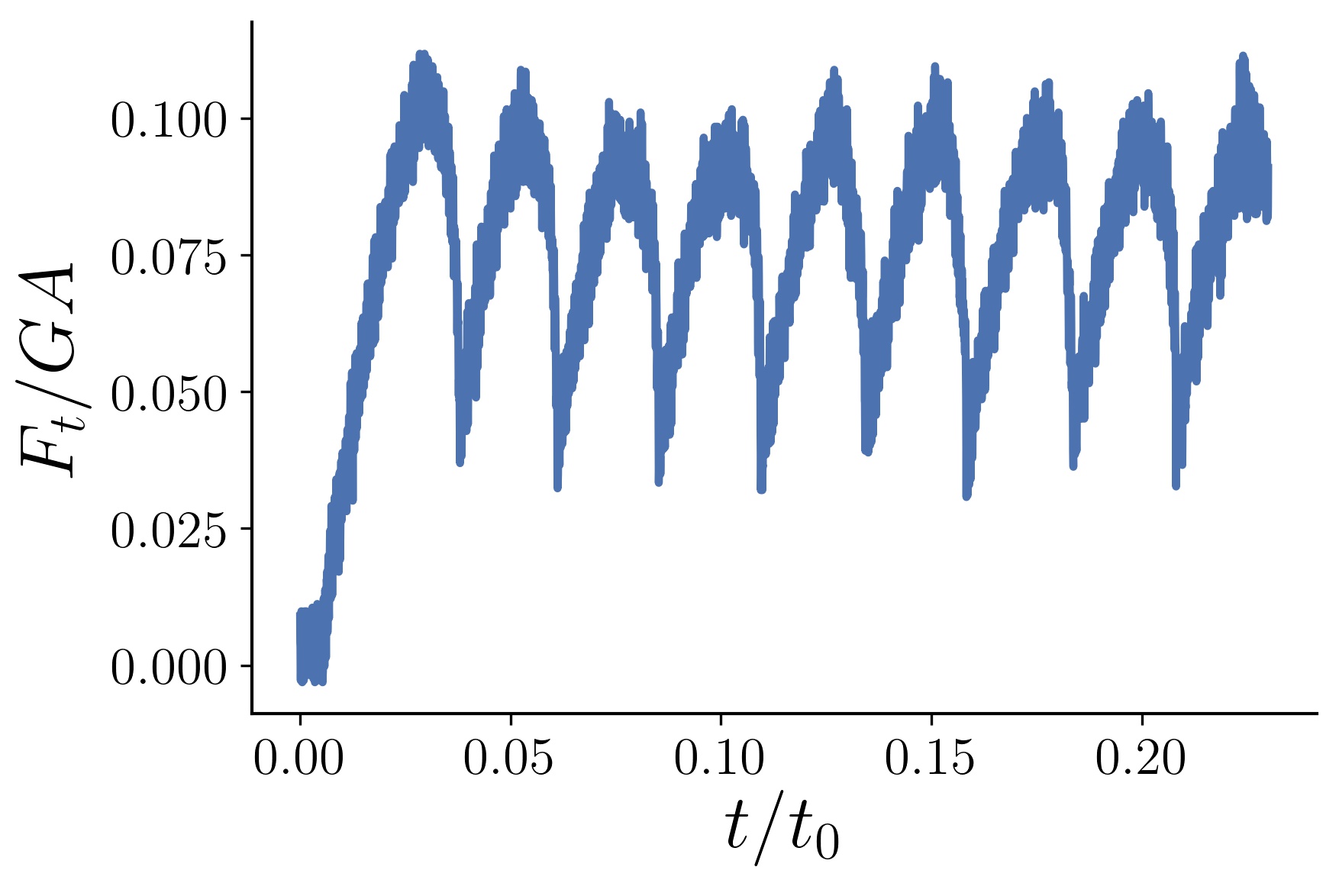}  
  \end{minipage}
  \begin{minipage}{0.48\textwidth}
    \includegraphics[width=\textwidth]{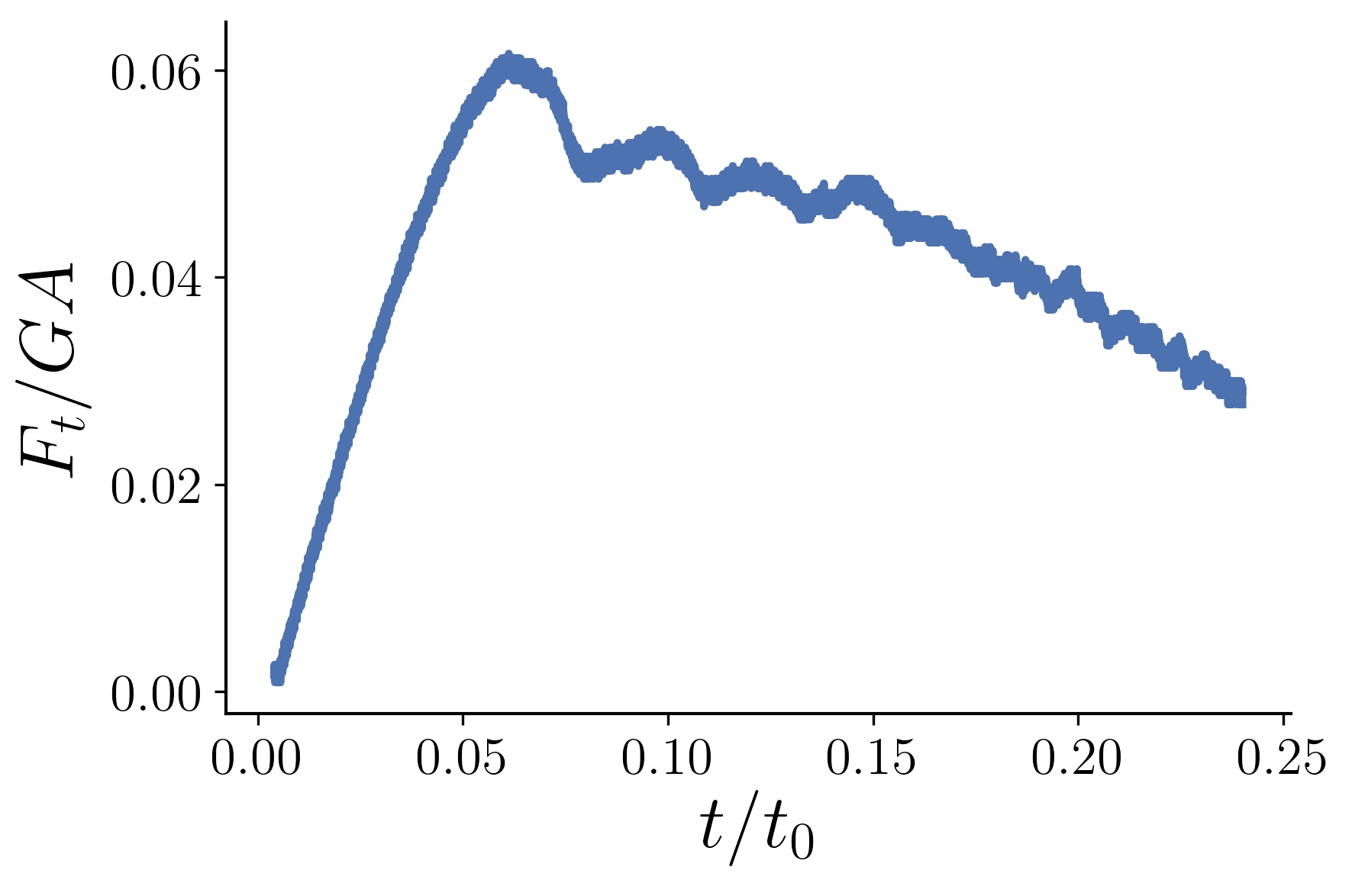}
  \end{minipage}
  \caption{Shear force measured simultaneous with multiple wave propagation events for Schallamach ($+$) waves (left) and separation pulse ($-$ wave, right). The horizontal axis is non-dimensionalized by time $t_0$ needed for the entire interface to slip by a unit distance. Applied velocity $\sV =$ 0.5 mm/s (left) and 0.05 mm/s (right) }%
  \label{fig:forces}
\end{figure}

At the macroscale, the effect of intermittent interface motion due to repeated wave propagation events is recorded by measuring the shear force as a function of time, see Fig.~\ref{fig:forces}. The corresponding non-dimensional force traces for $+$ and $-$ waves are shown in the left and right panels of this figure, respectively. The shear force is normalized by the product $GA$ of the shear modulus $G$ and the nominal contact area $A$. Time is non-dimensionalized by the time taken $t_0$ for the entire length $L$ of the interface to slip uniformly at speed $\sV$. The $\sV$ values for the $+$ and $-$ waves corresponding to the data in this figure are 0.5 mm/s and 0.05 mm/s, respectively.

Repeated wave motion results in oscillatory shear forces---a characteristic of stick--slip motion---with distinct frequency and amplitude reduction. For the case of $+$ waves, Fig.~\ref{fig:forces}(left), the force initially builds up as the interface is stationary and the shear stress on the interface increases. This built up stress is then released by the propagation of a single $+$ wave and a corresponding force reduction is observed in the figure. Each subsequent cycle corresponds to the propagation of one such wave at the interface. The time-scale for wave propagation is also much smaller than that for remote sliding $t_0 = L/\sV$. In comparison with $+$ waves, the propagation of $-$ waves is accompanied by a smaller force reduction and a larger frequency $n_{-}$. Furthermore, the force trace eventually decays to zero after the first few wave events likely due viscoelastic relaxation.

In summary, interface motion is not continuous but occurs intermittently via unit slip events. Depending on the remote $\sV$ and the applied normal load, these slip events are either mediated by Schallamach $(+)$ or separation $(-)$ waves that propagate parallel or anti-parallel to $\sV$, respectively. Both waves propagate at speeds $c_+, c_{-}$ that are much slower than any characteristic elastic wave speed in the material. They also retain their shape within the interface and result in a unit slip $\Delta x$ in the same direction as $\sV$. The net macroscale effect of this wave propagation is stick--slip motion of the interface. We reiterate that we have not discussed any of the complexities involved with wave nucleation, these have been reported for $+$ waves in an earlier manuscript \cite{ViswanathanETAL_PhysRevE_2015}. 

\section{Mechanics of detachment wave propagation}
\label{sec:mechanics}
We now attempt to explain the propagation of $\pm$ waves using a 2D version of the linear elastic framework introduced in Ref.~\cite{ViswanathanETAL_SoftMatter_2016_2}. Firstly, we assume that the elastic polymer and rigid indenter occupy the $z> 0$ and $z \leq 0$ half-spaces, respectively, see Fig.~\ref{fig:exptSchematic}, panel (b). Remote sliding $\sV$ is applied far away from the $z=0$ interface with a wave propagating at speed $c$ on $z=0$. We allow for wave propagation in both $+x$ and $-x$ directions; the length of the domain is $L$. In order to eliminate uncertainites induced by wave nucleation (initial conditions), we use periodic boundary conditions at $x = \pm L/2$. Furthermore, since detachment waves propagate at constant speed, we work in a co-moving frame of reference $\eta = k(x - ct)$ in which the wave is stationary. Next, the wave itself is constituted by a detached zone of extent $2\alpha$ in co-moving angular coordinates, which we take to be equal to the amount of slip induced $\Delta x$ (\emph{cf.} Figs.~\ref{fig:spaceTimeScW},~\ref{fig:spaceTimeSP}).

Without loss of generality, we set the detached zone as $-\alpha < \eta < \alpha$, with the rest of the interface ($\alpha<|\eta|< \pi$) in sticking contact. For this configuration, the boundary conditions on the interface $z=0$ are written as:
\begin{equation}
  \label{eqn:BCs}
  \sigma, \tau
  \begin{cases}
    =0 &\eta \in (-\alpha,\alpha)\\
    \neq 0 & \alpha < |\eta| < \pi
  \end{cases}
  \quad\quad\quad
  \dot{u}_x,\dot{u}_z
  \begin{cases}
    \neq 0 & \eta \in (-\alpha, \alpha)\\
    =0 & \alpha < |\eta|< \pi
  \end{cases}
\end{equation}
where $\sigma$ and $\tau$ are the normal and shear tractions on $z=0$. The far field normal and shear stresses are denoted $\sigma_r$ and $\tau_r$, respectively and the remote sliding speed is $\sV$. Note that only one of either $\tau_r$ or $\sV$ can be specified, the other is a response; in our case $\sV$ is applied,as in the experiments.

%To keep $\tau_r  > 0$, we assume $\sV$ along $x<0$ applied to the elastic polymer.

Once the wave completely passes any point within the interface, it causes that point to slip by a unit distance $\Delta x$. Consequently, after the wave has completely passed from $x = -L/2$ to $x = +L/2$, the entire interface has slipped by $\Delta x$ so that
\begin{equation}
  \label{eqn:slipBC}
  u_x =
  \begin{cases}
    0 \quad &\text{before wave}\\
    \Delta x \quad &\text{after wave passage}
  \end{cases}
  \quad\quad\quad \frac{\dot{u}_x(\eta)}{\sV} \geq 0 \quad \forall \,\eta \in (-\pi, \pi)
\end{equation}
The second condition on $\dot{u}_x$ arises from the fact that the interface always moves in the direction of imposed $\sV$ irrespective of wave motion direction, \emph{cf.} Fig.~\ref{fig:dWaveFrames}. 

Using physical arguments, it can be established \cite{ViswanathanETAL_SoftMatter_2016_1} that the wave parameters $\alpha, k, c$ can be related to the experimental parameters $\sV, \Delta x$ as
\begin{equation}
  \alpha =\pi \left|\frac{\sV}{c}\right|
  \quad\quad k=\frac{2\pi }{\Delta x}\left|\frac{\sV}{c}\right|
\end{equation}
so that the entire problem is posed in terms of the imposed loading $\sV$, and the observed parameters $c, \Delta x$.

Following the same dual series expansion procedure as outlined in Ref.~\cite{ViswanathanETAL_SoftMatter_2016_2} and exploiting the fact that the wave speeds are much smaller than both the longitudinal $(c_l)$ and transverse $(c_t)$ elastic wave speeds $c/c_l, c/c_t \ll 1$, we obtain the governing equations for the interface velocities $\dot{u}_x, \dot{u}_z$ in the form of coupled singular integral equations (SIEs):
\begin{equation}
  \label{eqn:governingSIE}
\begin{aligned}
 0&=\frac{\tau_r}{G}+\frac{2k_1}{\pi}\Big((1+a^2p^2)\int_{-1}^1\frac{ \Psi(s)ds}{s-p}+\pi a \frac{\sV}{c} p \Big)-2k_2(1+a^2p^2)\Phi(p)
\\
 0&=\frac{\sigma_r}{G}-2k_2\frac{\sV}{c}+\frac{2k_1}{\pi}(1+a^2p^2)\int_{-1}^1 \frac{\Phi(s)ds}{s-p}+2k_2(1+a^2p^2) \Psi(p)
 \end{aligned}
\end{equation}
Details of this derivation are provided as supplementary material. The material constant $G$ denotes the shear modulus and $a=\tan \alpha/2$. The angular variable $\eta$ is changed to $p=\frac{\tan \eta/2}{\tan \alpha/2}$ and the dimensionless constants $k_1=\frac{2(1-\nu)}{3-4\nu},\; k_2=\frac{1-2\nu}{3-4\nu}$ where $\nu$ is the elastic material's Poisson ratio. Finally, the unknowns $\dot{u}_x$ and $\dot{u}_z$ are expressed in the form of non-dimensional functions $\Psi(s)=\frac{\dot{u}_x(s)}{c(1+a^2s^2)}$ and $\Phi(s)=\frac{\dot{u}_z(s)}{c(1+a^2s^2)}$, respectively.

The SIEs in Eq.~\ref{eqn:governingSIE} have to be solved for $\Phi$ and $\Psi$, subject to the additional conditions
\begin{equation}
  \label{eqn:sideCondition}
  \int_{-1}^1\Psi(s)ds=\frac{\pi \sV}{a c} \quad\quad \int_{-1}^1\Phi(s)ds=0
\end{equation}
which result from orthogonality of the corresponding Fourier expansions \cite{ViswanathanETAL_SoftMatter_2016_2}. The SIEs in Eq.~\ref{eqn:governingSIE} are coupled and of the second-kind, making their solution analytically cumbersome. However, in principle, solving this system for $\Phi, \Psi$ gives us the interface velocities directly. Interface stresses $\sigma(\eta), \tau(\eta)$ are obtained using the RHS of Eq.~\ref{eqn:governingSIE}, but for $\alpha < |\eta| < \pi$. Note that the stresses are non-zero only for $\alpha < |\eta| < \pi$ (or $|p| \geq 1$) and the velocities for $|\eta| < \alpha$ (or $|p| < 1$). Interface displacements may be found by directly integrating the velocities $\dot{u}_x, \dot{u}_z$.

\subsection{Exact solution for $\nu = 0.5$: Existence of $+$ and $-$ waves}
\label{subsec:exactSoln}

When the elastic polymer is incompressible ($\nu=0.5$), the constants $k_1 = 1, k_2 = 0$ in Eq.~\ref{eqn:governingSIE} so that the two SIEs are decoupled. In effect, this implies that the normal and tangential direction stresses are completely independent, and makes the problem analytically tractable. The corresponding interface velocities are found by inverting the resulting uncoupled SIEs using standard techniques \cite{Barber_2002}
\begin{align}
  \label{eqn:incomp_velocities}
  \Phi(p)&=- \frac{\sigma_r}{2G}\frac{p}{\sqrt{1-p^2}}\frac{\sqrt{1+a^2}}{1+a^2p^2}\\\notag
  \Psi(p)&=\frac{1}{\sqrt{1-p^2}}\left[\left(\frac{\sV}{c\,a}\right)\frac{\sqrt{1+a^2}}{1+a^2p^2}-\frac{\tau_r}{2G}\frac{p\sqrt{1+a^2}}{1+a^2p^2} \right]
\end{align}
from which the final dimensional forms of $\dot{u}_x(x,t), \dot{u}_z(x,t)$ are easily obtained.

The interface stresses $\sigma(x,t)$ and $\tau(x,t)$ are similarly evaluated using the RHS of Eq.~\ref{eqn:governingSIE} for $\alpha < |\eta| < \pi$: 
\begin{align}
  \label{eqn:incomp_stresses}
  \frac{\sigma}{\sigma_r}&=\frac{|p|\sqrt{1+a^2}}{\sqrt{p^2-1}}\\\notag
  \frac{\tau}{\tau_r}&=\frac{|p|\sqrt{1+a^2}}{\sqrt{p^2-1}}-\left(\frac{\sV}{ac}\right)\left(\frac{2G}{\tau_r}\right)\frac{\sqrt{1+a^2}}{\sqrt{p^2-1}}\frac{|p|}{p}
\end{align}

Enforcing the $\dot{u}_x(\eta)$ constraint Eq.~\ref{eqn:slipBC} on the expression obtained from Eq.~\ref{eqn:incomp_velocities} results in the condition
\begin{equation}
  \label{eqn:incomp_bifDig}
  \left(\frac{\tan(\pi \sV/2c)}{\pi \sV/2c}\right) \geq \frac{4G}{\pi \tau_r}
\end{equation}

\begin{figure}
  \centering
  \includegraphics[width=0.65\textwidth]{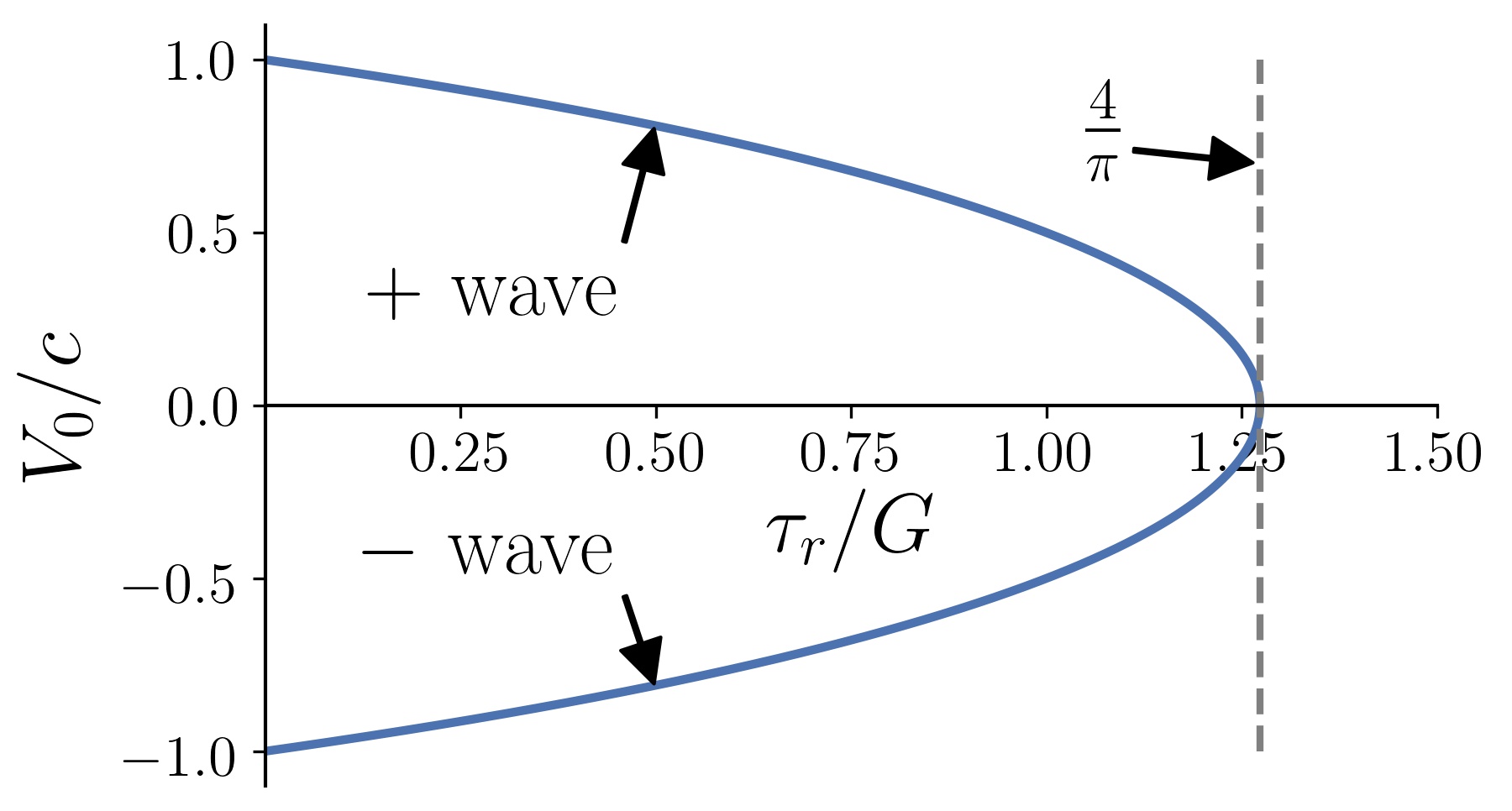}
  \caption{Diagram showing existence of two branches corresponding to $\pm$ detachment waves. Any $c$ between the two bounding curves is permitted for wave propagation. The limiting shear stress $\tau_r/G = 4/\pi$ is shown as a dashed line and represents the formal limit beyond which wave propagation ceases. }
  \label{fig:bifDig}
\end{figure}

This inequality, involving $\sV, c$ and $\tau_r$ puts a constraint on the existence of physically relevant wave solutions to the governing SIEs. Consequently, we may plot an existence diagram using Eq.~\ref{eqn:incomp_bifDig}, see Fig.~\ref{fig:bifDig}. The diagram has several important features. Firstly, it shows two branches satisfying Eq.~\ref{eqn:incomp_bifDig}; one for $c/\sV >0$ ($+$ wave branch) and the other for $c/\sV < 0$ ($-$ wave branch). Secondly, the minimum allowable wave speed $c_+$ or $c_{-}$ is set by the bounding blue curve; all wave velocities above this threshold (and therefore lying between the two curves) are permissible. Thus, the wave speed is not fixed \emph{a priori} by a material wave-speed but is instead determined by the interface conditions post wave nucleation. Thirdly, the $+$ wave and $-$ wave branches are perfectly symmetric and they merge at $\tau_r/G = 4/\pi$. At this point, the envelope of allowed wave speeds shrinks to zero and for $\tau_r/G > 4/\pi$, physically relevant propagating wave solutions are no longer possible. This diagram hence shows the domain of existence of opposite moving slow $\pm$ detachment waves within the elastic framework.

The interface displacements are obtained for both branches in Fig.~\ref{fig:bifDig} by integrating the velocities in Eq.~\ref{eqn:incomp_velocities}. The vertical displacement in the detachment zone is determined from $\dot{u}_z$
\begin{equation}
  \label{eqn:incomp_uz}
\frac{{u}_z}{\Delta x}= -\left(\frac{c}{2\pi\sV}\right)\left(\frac{\sigma_r}{G}\right)\tanh^{-1} \Big(\frac{a\sqrt{1-p^2}}{\sqrt{1+a^2}} \Big)
\end{equation}
Note that the constant of integration is fixed by the condition that the interface readheres after wave passage and no net $z$-displacement is induced for both $\pm$ waves.

The $x$-displacement is also obtained similarly, but the constant of integration must now be set a little more carefully depending on which branch in Fig.~\ref{fig:bifDig} we are describing. This is because the location of points before and after wave passage depends on the sign of $c/\sV$. Hence, the condition on $u_x$ from Eq.~\ref{eqn:slipBC} must be enforced separately for the $+$ and $-$ branches.

For a Schallamach ($+$) wave, we obtain
\begin{equation}
  \label{eqn:incomp_disp_x_P}
  \frac{{u}_x}{\Delta x}=-\frac{1}{\pi}\left[\tan^{-1}\left(\frac{p\sqrt{1+a^2}}{\sqrt{1-p^2}} \right) +\frac{c\tau_r}{2 G \sV }\tanh^{-1} \left(\frac{a\sqrt{1-p^2}}{\sqrt{1+a^2}} \right)\right]+\frac{1}{2}
\end{equation}
while for the separation ($-$) wave,
\begin{equation}
  \label{eqn:incomp_disp_x_M}
\frac{{u}_x}{\Delta x}= \frac{1}{\pi}\left[\tan^{-1}\left(\frac{p\sqrt{1+a^2}}{\sqrt{1-p^2}} \right) -\frac{c\tau_r}{2 G \sV }\tanh^{-1} \left(\frac{a\sqrt{1-p^2}}{\sqrt{1+a^2}} \right)\right]+\frac{1}{2}
\end{equation}

These relations fully resolve the interface dynamics for $\nu = 0.5$. Features of the solutions in Eqs.~\ref{eqn:incomp_velocities},~\ref{eqn:incomp_stresses}, ~\ref{eqn:incomp_disp_x_P},~\ref{eqn:incomp_disp_x_M} are presented in the next section after discussing the solution of the fully coupled ($\nu \neq 0.5$) problem.

%It is interesting to note that both waves can be modelled under the same framework and give out the same expressions for all elastic fields \footnote{Under the transformation $c$ $\rightarrow$ $-c$} except for the horizontal displacement.This is due to the fact that irrespective of which direction the wave goes, the local material should always move along the far field velocity. We explore this more in the next section.

\subsection{Interface dynamics for general elastic polymers $\nu \neq 0.5$}

For general elastic media with arbitrary $\nu$, the original SIE system Eq.~\ref{eqn:governingSIE} can be cast in the form of a matrix equation:
\begin{equation}
  \label{eqn:governingSIE_matrix}
 \begin{bmatrix}
-m & 0\\
0 & m
\end{bmatrix} \begin{bmatrix}
\Phi (p)\\
\Psi  (p)
\end{bmatrix} 
+\frac{1}{\pi} \begin{bmatrix}
0 & 1\\
1 & 0
\end{bmatrix}  \int_{-1}^1 \begin{bmatrix}
\Phi (u)\\
\Psi  (u)
\end{bmatrix} \frac{du}{u-p}=\begin{bmatrix}
-\frac{\tau_r}{2k_1 G (1+a^2p^2)}-2a\sV/c\\
\frac{2k_2\sV/c-\sigma_r/G}{2k_1 (1+a^2p^2 )}  
\end{bmatrix} 
\end{equation}
where $m=k_2/k_1$ and with the additional conditions of Eq.~\ref{eqn:sideCondition} as before.

In order to solve this coupled system numerically, we invert the SIEs using special function approximations \cite{ErdoganETAL_1973}. First, we introduce the complex variable $\chi= \Phi+i\Psi$  so that Eq.~\ref{eqn:governingSIE_matrix} is
\begin{equation}
  \label{eqn:governingSIE_complex}
 -m \chi  -\frac{i}{\pi}\int_{-1}^1 \frac{\chi(u)du}{u-p}=g_1-ig_2=g(\chi)
\end{equation}
Where the RHS is represented as $(g_1 \quad g_2)^T$ without loss of any generality. This type of equation can now be solved approximately using Jacobi polynomials $P_n^{(\xi,\zeta)}$. Just as with the $\nu=0.5$ case, we expect the solution to be singular at both ends, $p=\pm 1$ so that the singularity is automatically handled by the polynomials. To find the solution of Eq \ref{eqn:governingSIE_complex}, we consequently have to determine coefficients $c_n$ such that
\begin{equation}
\label{jacobi_exp}
\chi(t)= \sum_0^\infty c_n w(t)P_n^{(\xi,\zeta)}(t)
\end{equation}
with Jacobi weight functions for polynomials unbounded at both $p = \pm 1$ are given by
\begin{equation}
w(p)=(1-p)^\xi(1+p)^\zeta  \quad;\quad \xi=-\frac{1}{2}-i\omega \quad;\quad \zeta=-\frac{1}{2}+i\omega \quad \omega=\frac{1}{2\pi}\log \Big(\frac{1+m}{1-m}\Big) 
\end{equation}
with $m = \frac{1-2\nu}{2(1 - \nu)}$, as defined in Eq.~\ref{eqn:governingSIE_complex}.

The side condition Eq.~\ref{eqn:sideCondition} is now $\int_{-1}^1\chi(t)dt=\frac{i\pi \sV}{ac}$. An approximate solution for $\chi$ is readily obtained if the infinite series in Eq.~\ref{jacobi_exp} is terminated with a finite number of terms. Using orthogonality of Jacobi polynomials, the coefficients obey
\begin{equation}
  \label{eqn:Coeff_chi}
\frac{i}{2\sin \pi \xi}\theta_k(-\xi,-\zeta)c_{1+k}=F_k
\end{equation}
for $k=0,1,2, \ldots$
\begin{equation}
F_k= \int_{-1}^1P_k^{(-\xi,-\zeta)}(x)\frac{g(x)dx}{w(x)}
\end{equation}
\begin{equation}
\theta_k(\xi,\zeta)=\frac{2^{\xi+\zeta+1}}{2k+\xi+\zeta+1}\frac{\Gamma(k+\xi+1)\Gamma(k+\zeta+1)}{k!\Gamma(k+\xi+\zeta+1)}
\end{equation}
\begin{equation}
\theta_0(\xi,\zeta)= \frac{i\pi \sV}{ac} = \frac{2^{\xi+\zeta+1}\Gamma(\xi+1)\Gamma(\zeta+1)}{\Gamma(\xi+\zeta+2)}
\end{equation}
We can now solve for $c_1, c_2, \ldots$ using Eq.~\ref{eqn:Coeff_chi} and the corresponding expressions for $\theta_k$ and $F_k$.

The result of this numerical procedure is an $n$-term expansion for the interface velocities $\dot{u}_z, \dot{u}_x$ in terms of Jacobi polynomials. It was found that when the detachment zone is small ($a \ll 1$), only two terms in the expansion were sufficient. This was verified in two ways---one by solving the $\nu=0.5$ case numerically and comparing with the exact result presented in Sec.~\ref{subsec:exactSoln} and the other by explicitly verifying that $|c_n|/c_2 \ll 1$ for $n > 2$. The $\nu=0.5$ solution was reproduced almost exactly by the numerical scheme with $n=2$

We now discuss the various features of the interface velocities and stresses accompanying wave propagation obtained using this scheme. The existence of two wave solutions even for $\nu\neq 0.5$ may be seen by analytic continuation of the $\nu=0.5$ solution for arbitrary $\nu$. Additionally, the two solutions still remain symmetric as before (Fig.~\ref{fig:bifDig}) as explicitly verified by the change $c \to -c$. This is not unexpected since $\nu \neq 0.5$ only couples the normal and shear components but does not in any way change the symmetry in the problem. The results presented next are all for $\alpha = \pi/10$ and $\tau_r = 5\sigma_r$, unless specified otherwise.

\begin{figure}
  \centering
  \includegraphics[width=0.65\textwidth]{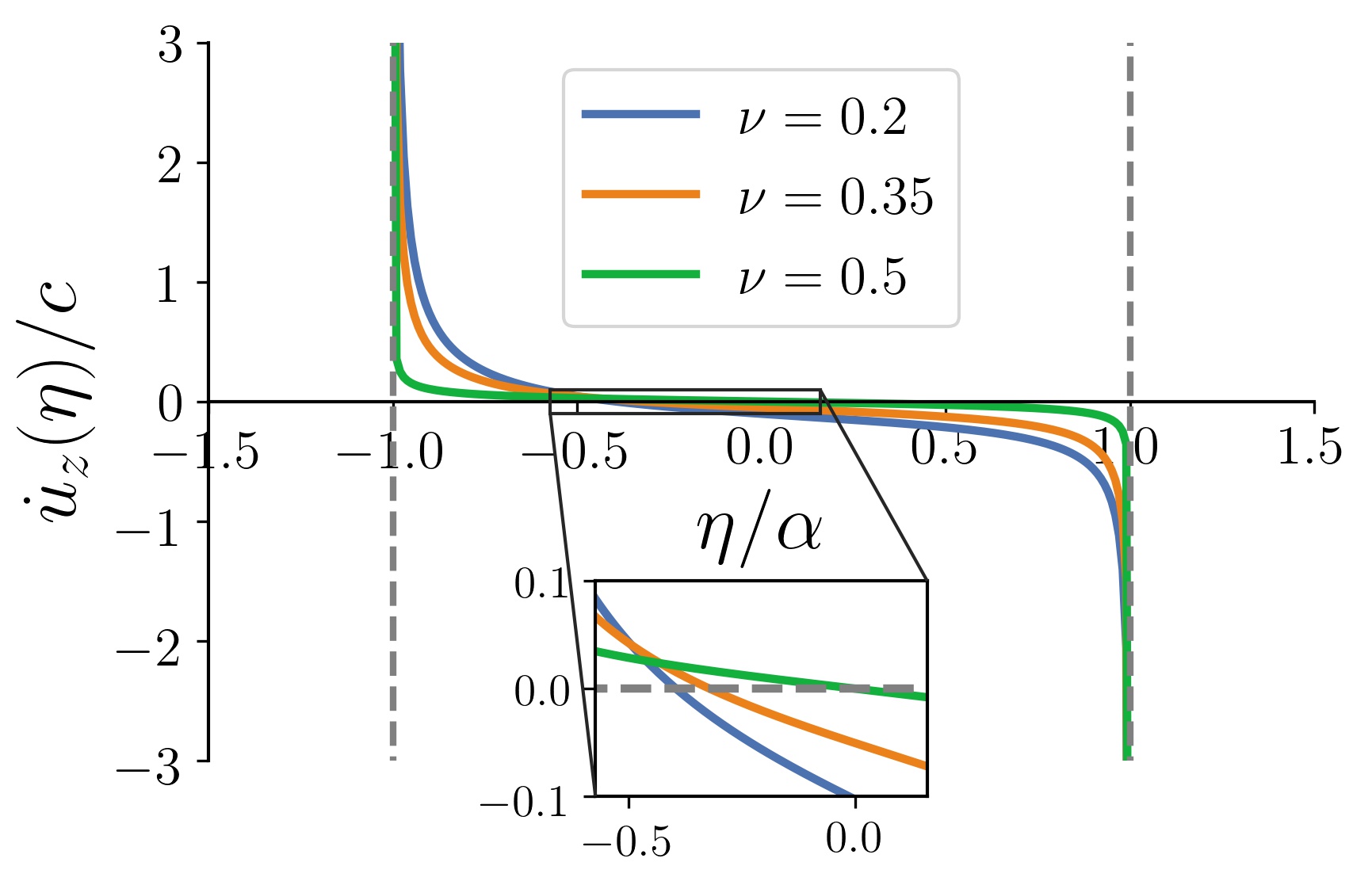}
  \caption{Interface $z$ velocity showing detaching and reattaching zones within the wave, as a function of $\nu$. Inset shows crossover points for $\dot{u}_z >0$ and $\dot{u}_z < 0$ for different $\nu$. }
  \label{fig:velZ}
\end{figure}

The vertical velocity within the detachment zone is shown in Fig.~\ref{fig:velZ} for $\nu = 0.2, 0.35, 0.5$ and $\alpha = \pi/10$. Note that the velocity is zero for $\alpha < |\eta| < \pi$ in accordance with the boundary conditions, Eq.~\ref{eqn:BCs}. Several features are immediately noticeable. Firstly, the positive (negative) part of the curve represents interface detachment (reattachment). The velocities are unbounded at the ends $\eta = \pm \alpha$ as expected from both the analytical and numerical solutions. Secondly, it is clear that the $\nu=0.5$ curve is perfectly antisymmetric with the interface stationary in the co-moving frame exactly at $\eta = 0$ (see inset). This stationary point is the same irrespective of both $\sV$ and $c$ ($+$ or $-$ waves), as can be checked from Eq.~\ref{eqn:incomp_velocities}. However, for $\nu = 0.2, 0.35$, the stationary point is away from $\eta = 0$, its precise location a function of $\sV, \sigma_r$ and $\tau_r$. As $\nu$ becomes smaller, the point shifts to the left (right) for $c>0$ $(<0)$. Finally, all three curves appear very close (but do not meet) near $\eta \simeq 0.3$. The only significance that may be attached with this point on the physical interface is that its velocity is nearly independent of $\nu$.

\begin{figure}
  \centering
  \includegraphics[width=0.65\textwidth]{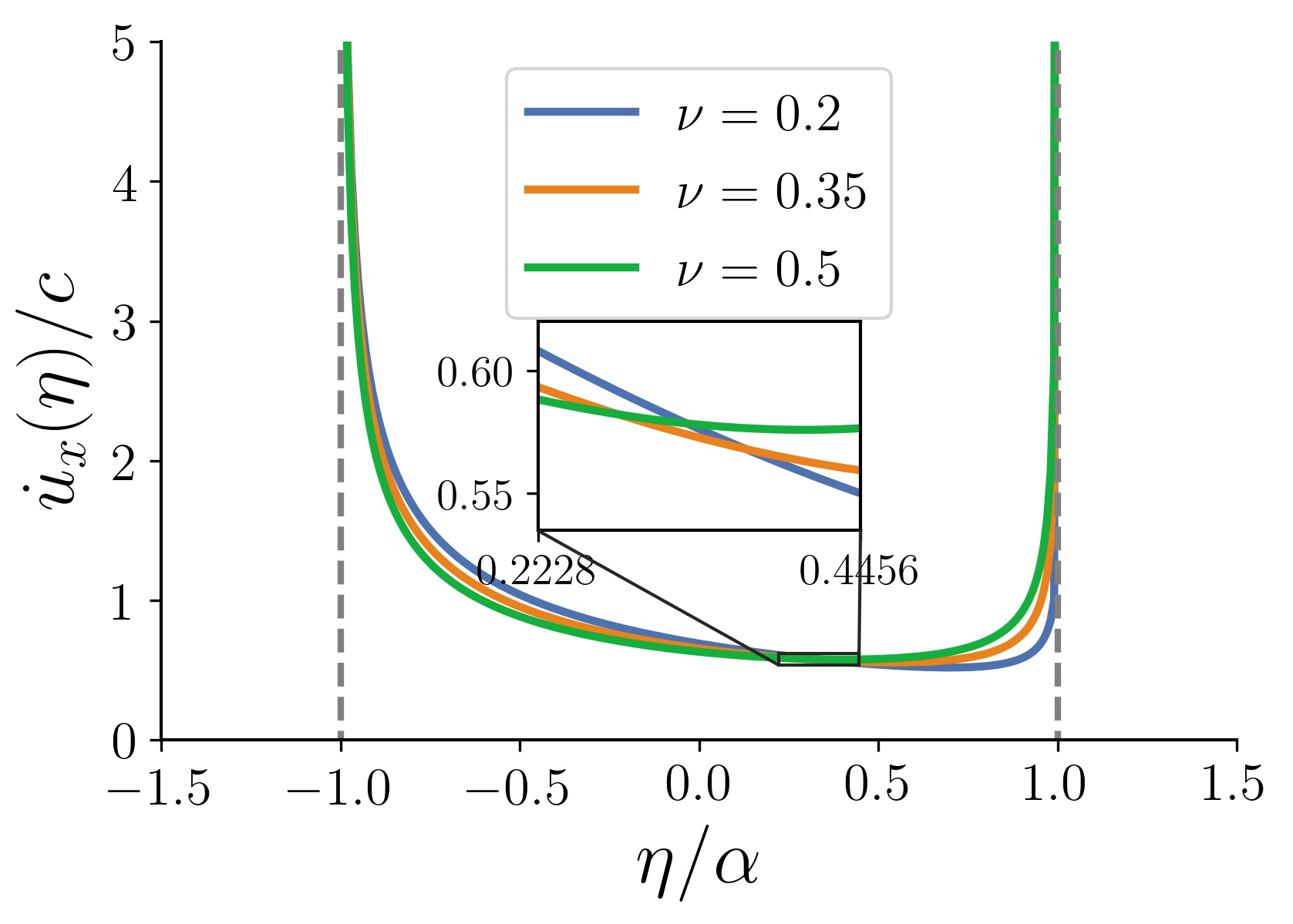}
  \caption{Interface horizontal velocity $\dot{u}_x(\eta)$  within the detachment zone for various values of $\nu$. Curves for different $\nu$ do not meet at a common point. }
  \label{fig:velX}
\end{figure}

The horizontal velocity shows very interesting dynamics, see Fig.~\ref{fig:velX}, for conditions identical to that for Fig.~\ref{fig:velZ}. Firstly, the velocity $\dot{u}_x/c$ is always positive as required by physical considerations, see Eq.~\ref{eqn:slipBC}. As with $\dot{u}_z$, it is also unbounded at both ends $\eta = \pm \alpha$ for all $\nu$ values. Secondly, the point $\eta \simeq 0.3\alpha$ again appears to be insensitive to $\nu$ (see inset) and the three curves for $\nu= 0.2,0.35, 0.5$ are very close to one another. This is expected, since at this point $\dot{u}_z$ is nearly independent of $\nu$ so that the coupling between the $x$ and $z$ directions is minimal. Finally, as $\nu$ deviates from 0.5, the trailing edge $\eta = -\alpha$  of the wave has a larger velocity gradient than the leading edge $\eta = +\alpha$.

\begin{figure}
  \centering
  \includegraphics[width=0.65\textwidth]{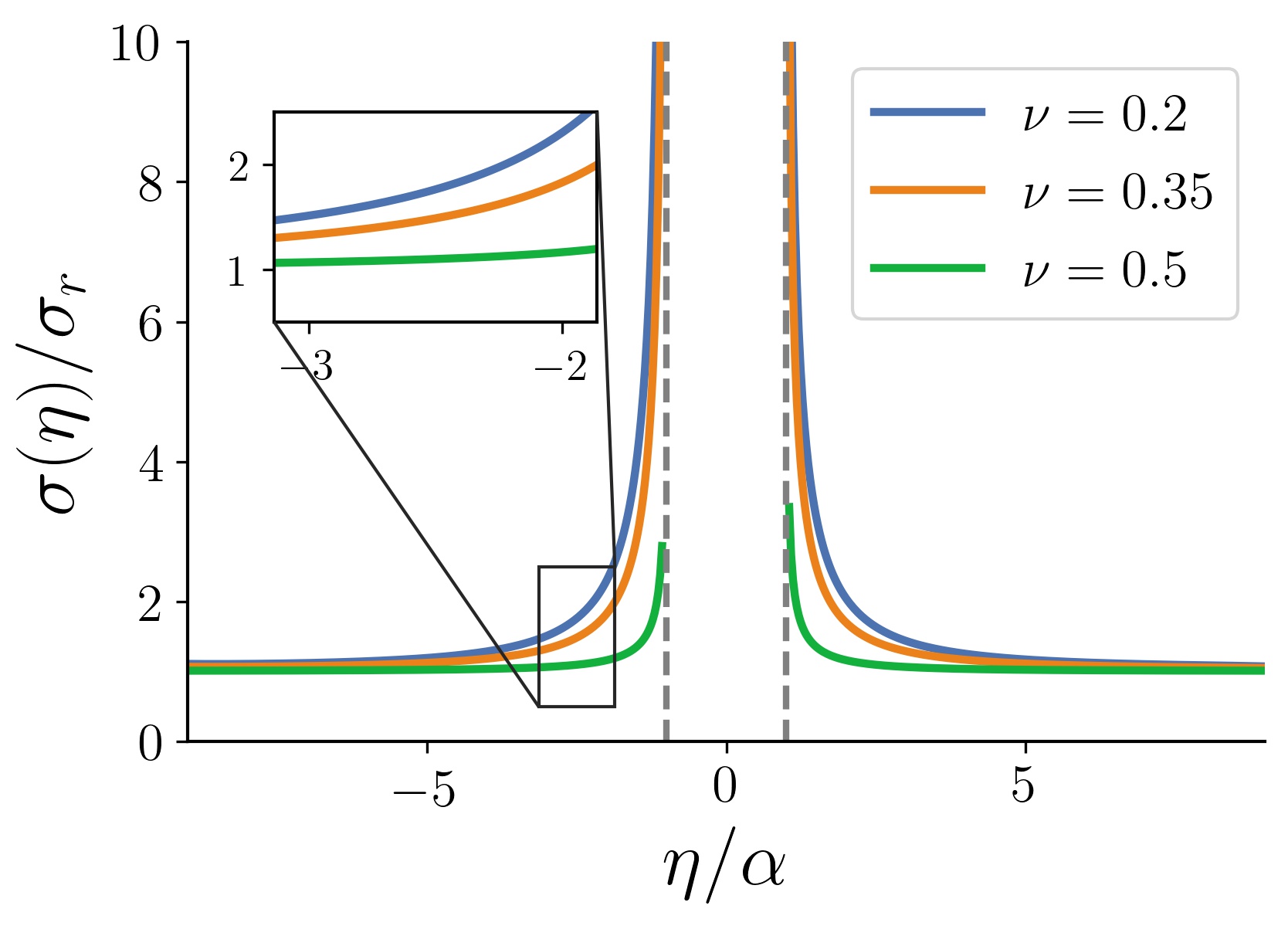}
  \caption{Normal stress variation along the interface ahead and behind a moving detachment wave. Note that the stresses are constant $\sigma(\eta) =\sigma_r$ far away from the detachment zone and zero within, as per the applied boundary conditions.}
  \label{fig:normSt}
\end{figure}

We now turn to the interface stresses as the wave propagates. These are complementary to the velocities and are non-zero only outside the contact zone $\alpha < |\eta| < \pi$. In the following, we assume that $\tau_r = 5\sigma_r$, in keeping with the experimental observations. The normal stress variation for $\nu = 0.2, 0.35, 0.5$ ($\alpha = \pi/10$ is fixed) is shown in Fig.~\ref{fig:normSt}. In all three cases, the normal stress is unbounded as $\eta \to \pm \alpha^{\pm}$. This is a well-known characteristic of adhesive contacts \cite{JohnsonETAL_ProcRoySocA_1971} and implies that need for adhesion emerges naturally from the solution of the problem, without explicit specification in any of the boundary conditions. Furthermore, the $\nu = 0.5$ curve immediately decays to the remote value $\sigma_r$ both ahead and behind the detachment zone. This decay is much more gradual for smaller $\nu$ so that significant normal stress deviation ($\sigma/\sigma_r \neq 1$) occurs over a larger part of the sticking zone when $\nu = 0.2$. This has important consequences for the occurrence of these waves, as discussed in Sec.~\ref{subsec:processZone}.

\begin{figure}
  \centering
  \includegraphics[width=0.65\textwidth]{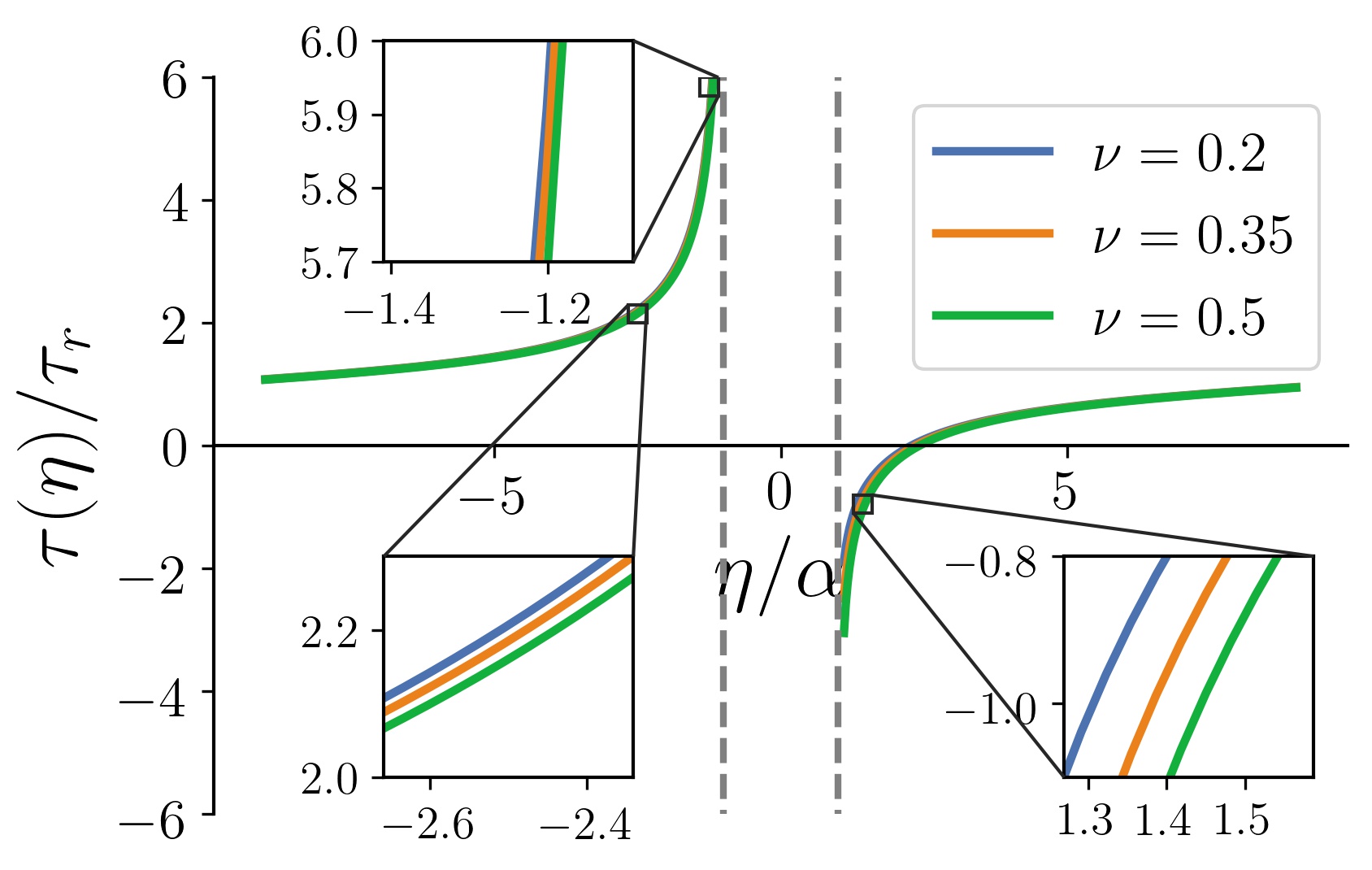}
  \caption{Tangential stresses accompanying the motion of a single detachment wave. The shear stress singularity changes sign on either side of the wave, corresponding to detachment and reattachment. $\tau/\tau_r \to 1$ as $\eta \to \pm \pi$ and $\tau/\tau_r = 0 $ within the detachment zone. }
  \label{fig:shearSt}
\end{figure}

The shear stresses for $\nu = 0.2, 0.35, 0.5$ are unbounded at $\eta \to \pm \alpha^{\pm}$ and appear to be nearly indistinguishable, see Fig.~\ref{fig:shearSt}. This is in contrast with the $\sigma$ curve and is due to $\tau_r$ being comparatively larger than $\sigma_r$ ($\tau_r = 5\sigma_r$ by assumption). They indeed differ very little near the $\eta = -\alpha$ edge or even for smaller $\eta < -\alpha$ (see insets). As $\eta \to \pi$, $\tau/\tau_r \to 1$ as is expected. However, the effect of $\nu$ is clearly seen in the zone $\eta > \alpha$. Here, the three curves are quite distinct (see inset, right) with the $\nu = 0.5$ curve decaying slowest, in contrast to the normal stress case (Fig.~\ref{fig:normSt}). Also most noteworthy are the opposite signs of the shear stress at the left and right ends of the detachment zone $\eta = \pm \alpha$. This is a consequence of the need for the interface to reattach under an applied remote shear load, its implications will be explored further in Sec.~\ref{subsec:toughening}. Finally, given this sign change, there exists a point on the interface where $\tau =0$, its exact location depends on $\nu$

\subsection{Slip due to $\pm$ waves}
\label{subsec:interfaceMotion}

\begin{figure}
  \centering
  \begin{minipage}{0.46\textwidth}
  \includegraphics[width=\textwidth]{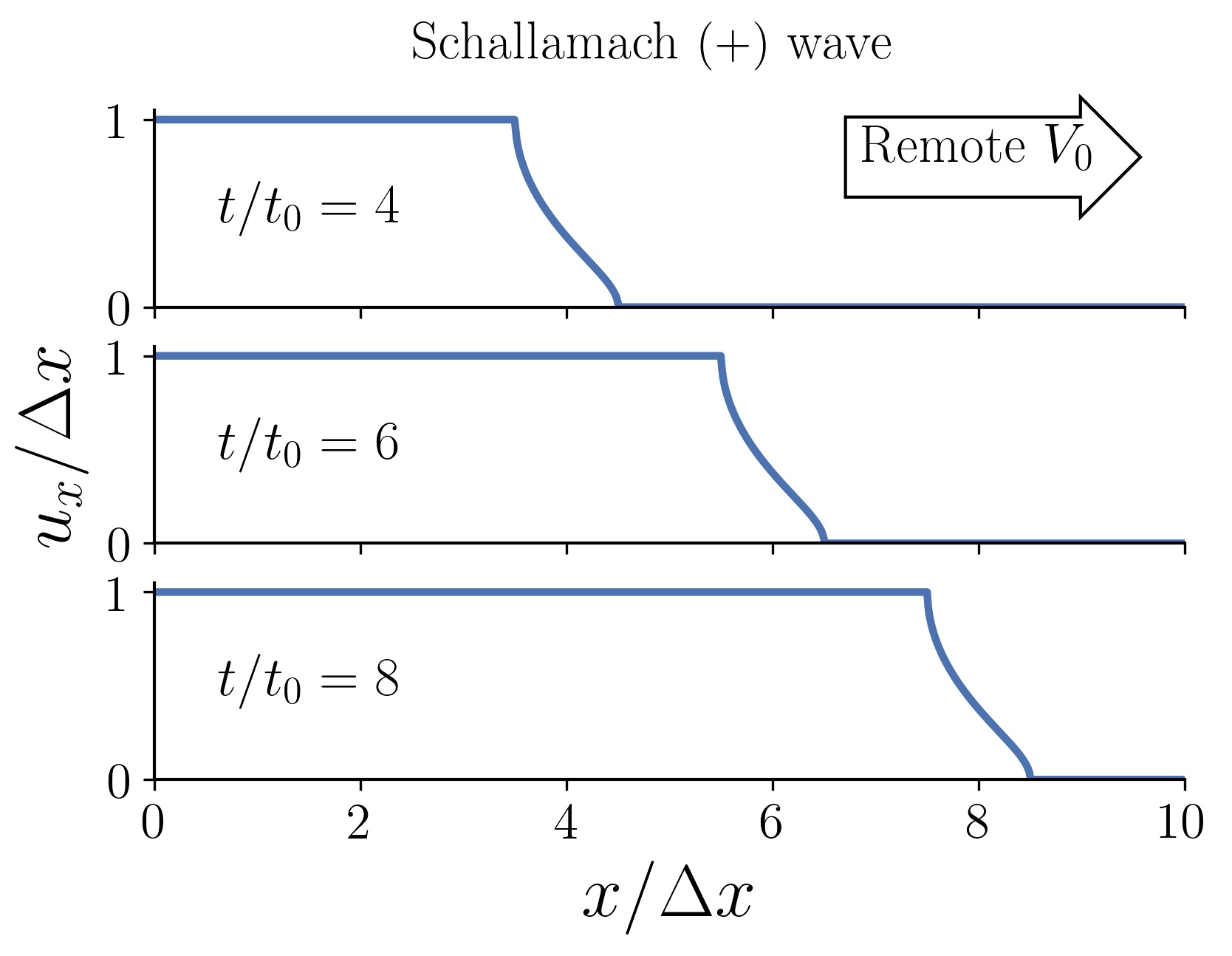}  
  \end{minipage}
  \qquad
  \begin{minipage}{0.46\textwidth}
    \includegraphics[width=\textwidth]{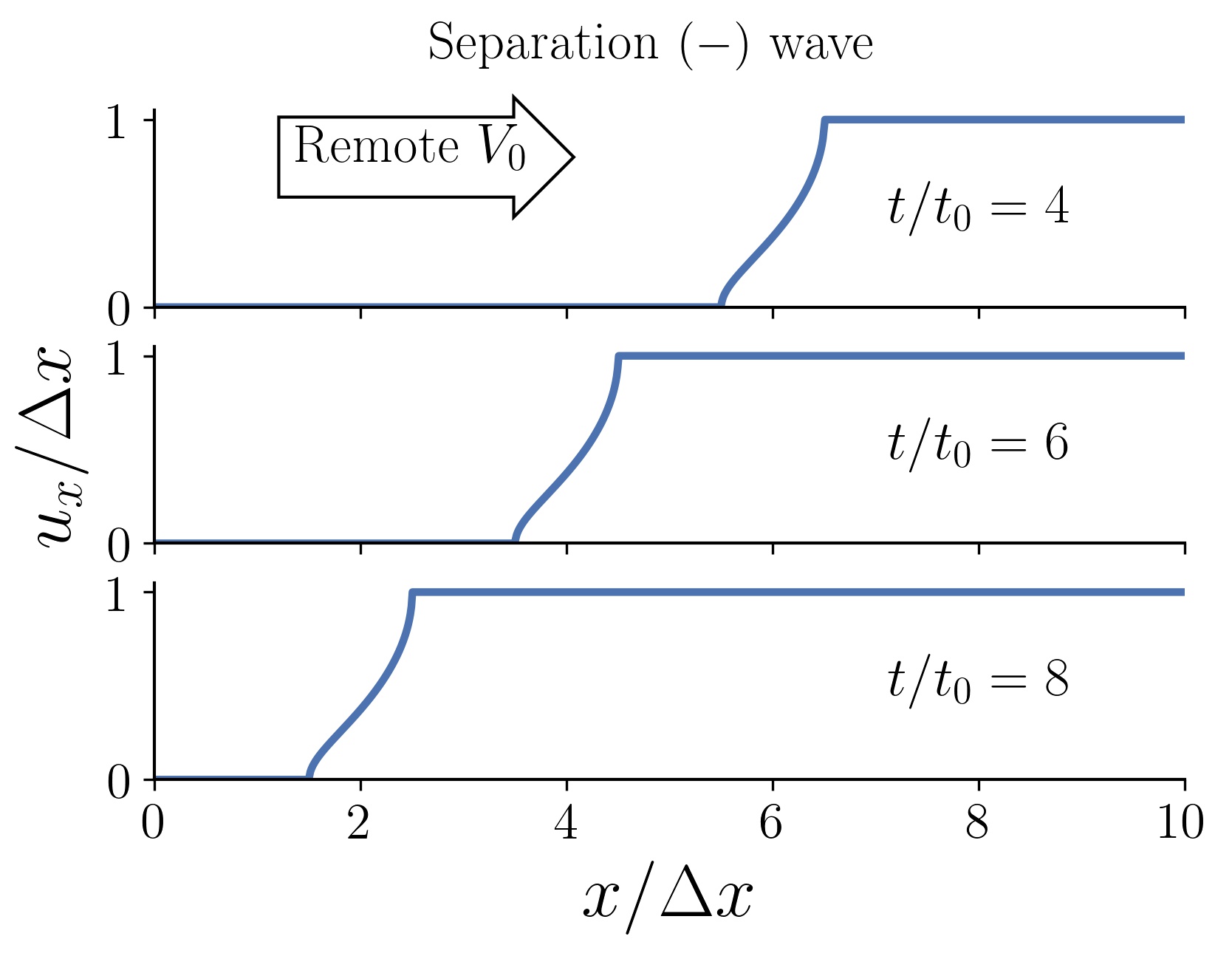}
  \end{minipage}
  \caption{Interface displacement due to passage of single $+$ (left) and $-$ (right) detachment waves. Note that the same solution is applicable to both types of waves so that interface slip $\Delta x$ is in the direction of $\sV$ Note that length in the horizontal direction is non-dimensionalized by $\Delta x$ and time $t$ by $t_0 = \Delta x/c$, i.e., the time taken for the wave to traverse a unit slip distance.}
  \label{fig:interfaceDisp}
\end{figure}

The displacements can be obtained from the interface velocities by time integration, with the resulting integration constant fixed according to the6 slip conditions in Eq.~\ref{eqn:slipBC}. We use the $\nu=0.5$ solution obtained in Sec.~\ref{subsec:exactSoln} to illustrate the nature of interface slip as a single $\pm$ wave propagates. A single-slip event with $u_x/\Delta x$ is shown in Fig.~\ref{fig:interfaceDisp} for both $+$ waves (left panel) and $-$ waves (right panel). The direction of remote $\sV$ is to the right in both panels. Note that $t$ is now non-dimensionalized by $t_0 = \Delta x/\sV$ as opposed to $L/\sV$.

For a $+$ wave, the wave propagation direction is the same as $\sV$ so that initially unslipped regions ($u_x/\Delta x = 0$) start progressively slipping from the left. This slip zone (separating slipped and unslipped regions) moves with the wave from left to right. The converse happens for the $-$ wave---the slip zone moves from right to left---even though the net overall slip $u_x/\Delta x$ changes again from $0$ to $+1$. Note that the displacement curves become vertical at either end of the detachment zone but remain bounded unlike the velocities (Fig.~\ref{fig:velZ},~\ref{fig:velX}). The next slip event proceeds analogously in either case so that a single wave may be thought of as a boundary separating slipped and unslipped regions of the interface.   
  
\subsection{Equivalence between $\pm$ waves and interface cracks}
\label{subsec:fractureEquivalence}

The elastic interface fields accompanying single $\pm$ detachment waves in Sec.~\ref{subsec:exactSoln}, especially the $1/\sqrt{r}$ stress singularities in Eq.~\ref{eqn:incomp_stresses} are strongly reminiscent of moving interface cracks. This analogy is actually exact in the limit $a \to 0$ and may be established for the $\nu=0.5$ case as follows. 

Firstly, the interface normal stress may be rewritten from Eq.~\ref{eqn:incomp_stresses} as
\begin{equation}
 \sigma=\frac{\sigma_r |\tan \eta/2|\sec\alpha/2 }{\sqrt{\tan^2 \eta/2-\tan^2 \alpha/2}}
\end{equation}
and is applicable for both $\pm$ waves by accounting for the sign of $c/\sV$.

As we approach the leading edge of the wave, $\eta=\alpha+\Delta \alpha$, with $\Delta\alpha \to 0^+$ so that the denominator may be factored using a Taylor series expansion in $\Delta \alpha$ as 
\begin{equation}
 \sigma =\frac{\sigma_r \tan (\alpha/2)\sec \alpha/2}{\sqrt{2\tan \alpha/2}\sqrt{\tan(\alpha/2+\Delta\alpha/2)-\tan \alpha/2}}\bigg|_{\Delta \alpha \to 0^+} = \frac{\sigma_r \sqrt{2\tan\alpha/2}}{\sqrt{k} \sqrt{r}}
\end{equation}
where $r$ is the distance (in physical coordinates) from the leading edge of the $\pm$ wave. Given this $1/\sqrt{r}$ dependence, we may define an equivalent mode-I stress intensity factor $K_{I}^{LE}$ for the leading edge as
\begin{equation}
 K_{I}^{LE}= \frac{\sigma_r\sqrt{2\pi \tan \alpha/2}}{\sqrt k}
\end{equation}
which, for $a \ll 1$, $\tan \alpha/2 \approx \alpha/2$ and $\alpha/k=\Delta x/2$, reduces to 
\begin{equation}
  \label{eqn:SIE_small_a}
K_I= \sigma_r\sqrt{\pi\Delta x/2}
\end{equation}
The normal stress is symmetric so that both the leading and trailing edges of the contact have the same $K_I$. Recall that this expression is obtained from the elastodynamic solution after taking $c/c_l, c/c_t \ll 1$ limit. This must be contrasted with a static fracture mechanics solution for an interface crack with a large mismatch in elastic moduli of two contacting incompressible materials \cite{RiceSih_JApplMech_1965}. The second Dundurs' parameter for this case is 0 so that the mode-I stress intensity factor for a crack of length $2l$ becomes identical with that in Eq.~\ref{eqn:SIE_small_a} if we replace $\Delta x$ by $2l$.

%For a geometry in which we have a large mismatch between the elastic moduli, i.e., $E_1/E_2 \gg 1$ and $\nu_2=0.5$, we have the second Dundur's parameter \textbf{ ref dundurs} as zero. If we insert this into the classical solution for a interface crack of width $2l$ between two elastic half spaces (\textbf{rice}) we exactly recover the SIF with $2l=\Delta x$.

An equivalent procedure can be repeated for the mode II shear loading, the only change being that we now have to account for sign changes at the trailing and leading edges. However, the magnitude of the leading edge stresses will remain the same for both $\pm$ waves. For $a\ll 1$, $\frac{2\sV}{a}\to\frac{4}{\pi}$ and an analogous calculation gives
\begin{equation}
K_{II}=\left(\tau_r+4/\pi \right)\sqrt{\pi\Delta x/2}
\end{equation}
This mode II stress intensity factor is again indentical with that for a static interface crack, provided we replace the remote shear $\tau_r$ by $\tau_r+4/\pi$ and the crack length $2l$ by $\Delta x$. The extra $4/\pi$ term arises from the remote velocity $\sV$ and has no analogue in the static case.

Both mode I and mode II stress concentration factors are functions of the interface slip $\Delta x$ which is equal to the detachment zone width. For $\pm$ waves that effect larger unit slip, the stress intensity at the leading edge is larger so that they may propagate at lower remote stress. However, it is important to remember that detachment wave propagation is a fundamentally different process on the macroscale compared to (catastrophic) crack growth. The former results in unit tangential slip at the interface, while the latter causes rupture and subsequent interface separation in the normal direction.

%What happens when $\nu \neq 0.5$? Again taking $E_2/E_1 \ll 1$, we obtain the second Dundurs parameter as $\frac{1-2\nu}{2(1-\nu)}$ for the plane strain case. This limit by itself is interesting because the term $m$ in the numerical scheme (Eq \ref{eqn_chi}) has the same representation in terms of Poisson ratio.

%Furthermore, our solution can be shown to be equivalent to an interface crack solution obtained via a completely different approach \textbf{ref}. The full static equivalent of the problem that we have solved is an array of interface cracks at the junction of two solid half spaces. 

% ===================================

\subsection{Bounded stresses and process zones for waves}
\label{subsec:processZone}

\begin{figure}
  \begin{minipage}{0.5\textwidth}
  \includegraphics[width=\textwidth]{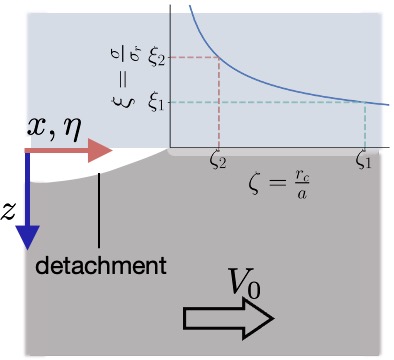} 
  \end{minipage}
  \begin{minipage}{0.5\textwidth}
    \includegraphics[width=\textwidth]{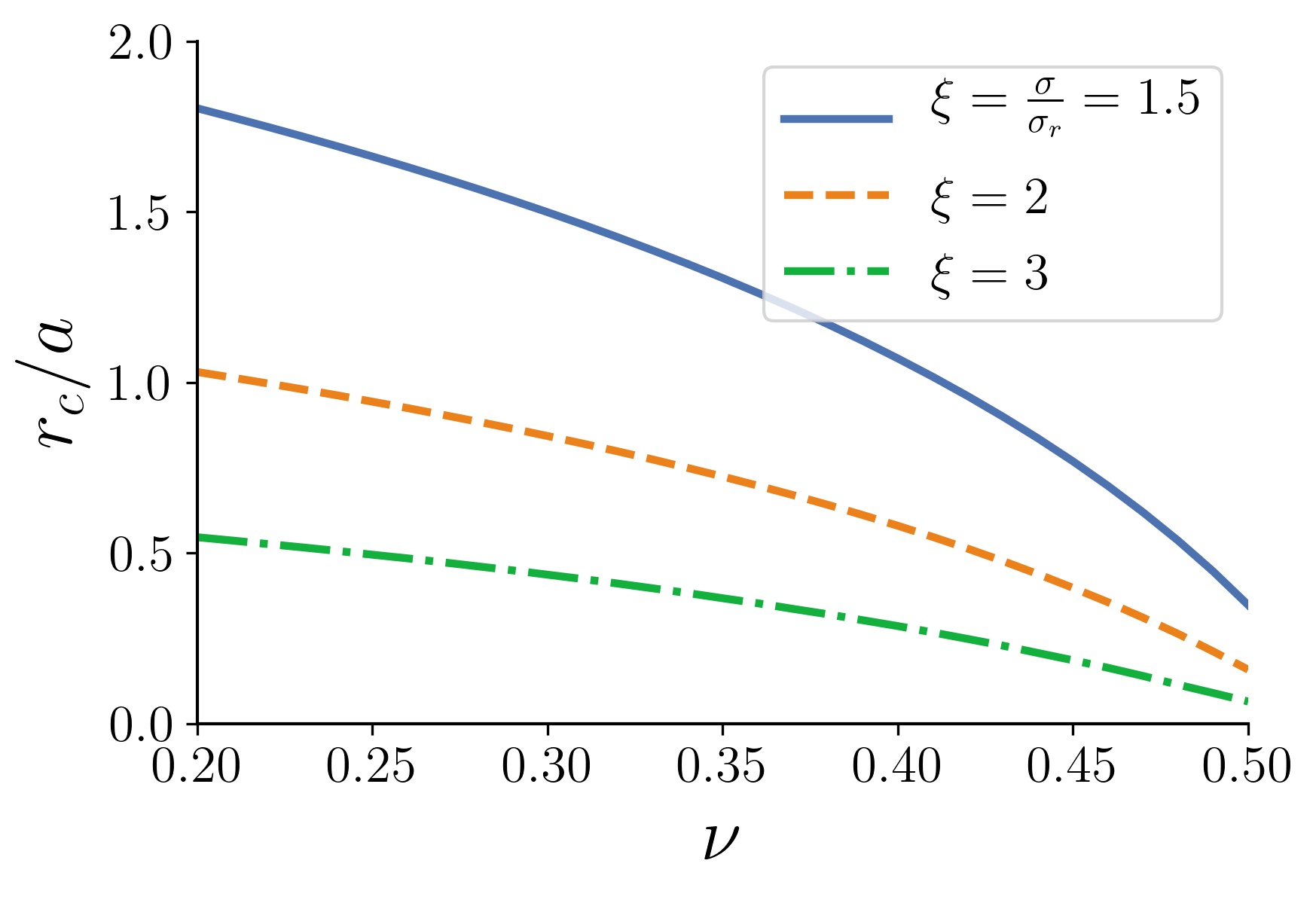}
  \end{minipage}
  \caption{Stresses ahead of the leading edge of a moving wave. (Left) Schematic of the \lq process zone\rq\ ahead of the wave representing the macroscopic effect of microscopic mechanisms that bound the stresses. (Right) Process zone size as a function of the Poisson ratio $\nu$, for different cut off values $\xi$.}
  \label{fig:processZone}
\end{figure}

Having established the equivalence between detachment wave propagation and interface cracks, we can use some fundamental results from interface fracture to make predictions about stick--slip. We first explore the consequences of a microscopic mechanism that bounds the stresses ahead of the leading edge of a detachment wave. This the analogue of plastic zone models employed in fracture of ductile materials \cite{BiblyETAL_ProcRoySocA_1963, Barenblatt_AdvApplMech}. Let the dimensionless cut-off stress be denoted by $\xi = \sigma/\sigma_r$. For various values of this cut-off threshold, we can evaluate the equivalent \lq plastic\rq\ or process zone size $r_c$ ahead of the leading edge, see Fig.~\ref{fig:processZone}. The schematic on the left shows the process zone and stress cut-off for either stress component. The panel on the right of this figure shows the variation of $r_c$ with $\nu$ for three different values of $\xi$, and has two interesting consequences that are noteworthy. Firstly, and quite understandably, $r_c$ decreases for any $\nu$ if the threshold $\xi$ is raised. This means that the material can sustain larger opening stresses prior to wave propagation. The precise microscopic mechanisms operating at the leading edge will set the exact value of $\xi$. Secondly, a large discrepancy in process zone size is seen for $\nu=0.2$, so that $r_c$ is sensitive to the chosen stress threshold $\xi$. This makes specification of the microscopic mechanisms bounding the stress important in systems where $\nu$ is small.

%Conversely, difference in $r_c$ becomes much smaller as we approach $\nu = 0.5$, likely indicating that $\nu=0.5$ is the most conducive for wave propagation.

\subsection{From stick--slip waves to steady sliding via stress intensity factors}
\label{subsec:toughening}

%The mode mixity is defined as $\vartheta=\tan ^{-1}K_2/K_1$, which reduces to $\tan^{-1}\frac{\tau_r}{\sigma_r}$ for $\nu=0.5$. It is well known that increasing mode mixity increases the effective fracture toughness of an interface crack \cite{LiechtiChai_JApplMech_1992}. 

\begin{figure}
  \centering
  \includegraphics[width=0.75\textwidth]{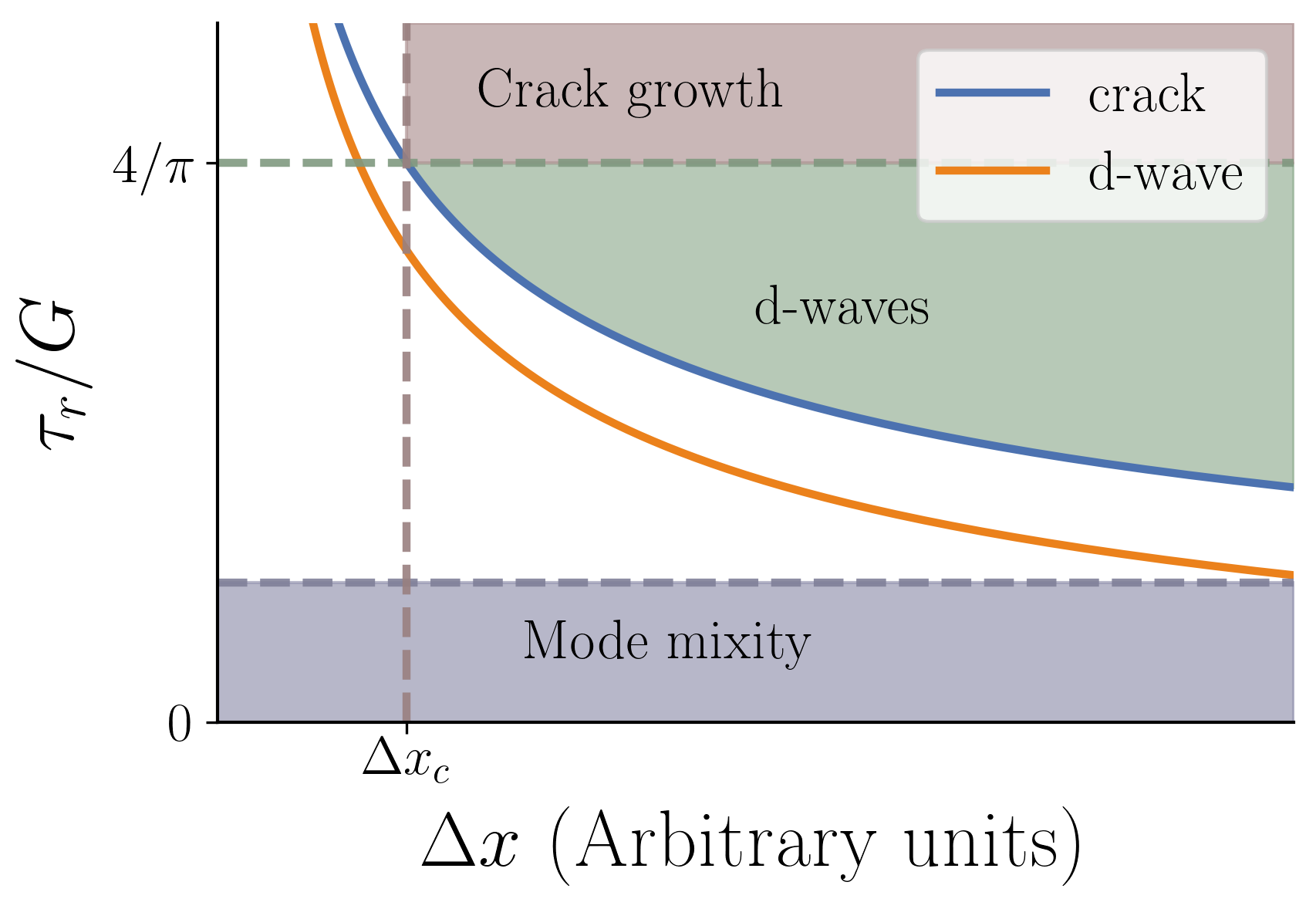}
  \caption{Phase diagram showing domains of occurrence of $\pm$ detachment waves (d-waves) and possible uniform steady sliding. The blue curve represents the stress intensity for shear interface cracks and the orange curve shows the corresponding intensity for detachment wave motion.}
  \label{fig:phaseDig}
\end{figure}

A second consequence of the fracture equivalence is that slow moving waves can potentially retard interface rupture or fracture under remote loading, leading to an apparent increase in interface toughness. Given that the onset of steady sliding at an elastic interface is via the propagation of a rupture front resembling a crack \cite{RubinsteinETAL_Nature_2004}, one of two sliding modes can occur: stick--slip (via detachment wave propagation) or steady sliding (via rupture or crack growth). To evaluate which one occurs for a given $\tau_r-\Delta x$ combination, we use the stress intensity factor at the leading edge of the wave, Eq.~\ref{eqn:SIE_small_a} and the corresponding value for a static interface crack. Firstly, for the leading edge of a detachment wave to advance, we require
\begin{equation}
  \label{pd1}
\Big(\tau_r+\frac{4G}{\pi}\Big)\sqrt{\pi \Delta x/2}=\Gamma
\end{equation}
In contrast, crack-like propagation at both ends requires
\begin{equation}
  \label{pd2}
\tau_r\sqrt{\pi \Delta x/2}=\Gamma
\end{equation}
We assume that the fracture toughness $\Gamma$ is independent of mode mixity $\vartheta$. Note that $\Gamma$ usually increases with $\vartheta$ but given that $\tau_r \gg \sigma_r$ from experiments, $\Gamma=$ constant is a reasonable assumption \cite{LiechtiChai_JApplMech_1992}. Secondly, $\Delta x$ is the equivalent crack length for both possible modes so that the geometry at the leading edge is identical. Finally, the detachment wave solution we have obtained is for the full domain and not just an asymptotic approximation as with static crack fields. So our inferences should apply to both types of $\pm$ waves, at the corresponding leading edge. We also know that the interface cannot sustain detachment waves for far field shear $\tau_r > \frac{4G}{\pi}$ (see Fig.~\ref{fig:bifDig}) which forms a boundary in $\tau-\Delta x$ space.

Based on these facts, we obtain the phase diagram shown in Fig.~\ref{fig:phaseDig}. The diagram is applicable to any material pair capable of forming adhesive contact as long as one material is significantly stiffer than the other. The primary geometric requirement is that the dimensions of the contacting solids must be much larger than the crack length in order to justify the half space assumption.

The diagram must be interpreted as follows. Consider an adhesive interface with an inherent detachment zone of size $\Delta x < \Delta x_C$. If such a material is loaded under remote $\sV$, crack propagation will begin at both ends of the detachment zone when the $\tau_r$ value reaches the blue curve. Given that the threshold $\frac{4G}{\pi}$ is not reached before the blue curve, this interface rupture will lead to complete slip, resulting in steady interface sliding. On the other hand, for $\Delta x > \Delta x_C$, rupture will again start when the remote shear reaches the corresponding value on the blue curve. However, now detachment wave propagation is possible at a lower $\tau_r$ so that sliding can occur via periodic detachment wave propagation, just as seen in Fig.~\ref{fig:forces}. Hence for all $\Delta x > \Delta x_C$, stick--slip via propagation of $\pm$ detachment waves is the likely mode of interface sliding. 

This entire picture places significant constraints on the nucleation of detachment zones at interfaces. For any material to show stick--slip motion consistently, it must be capable of producing $\Delta x$, via either buckling or tensile necking, that is larger than $\Delta x_C$. The propensity for producing such a large detachment zone is also likely the reason why polymers readily show stick--slip via detachment wave propagation.

\section{Discussion}
\label{sec:discussion}

Our results have shown that in soft adhesive interfaces, local interface motion via stick--slip occurs due to the propagation of detachment waves. These waves come in $+$ and $-$ varieties and cause effective displacement or slip in the same direction as the remote applied $\sV$. Wave speeds are constant within the interface, as evidenced by the space-time diagrams, see Figs.~\ref{fig:spaceTimeScW},~\ref{fig:spaceTimeSP}. Each stick-slip event results from a single wave moving through the interface; corresponding shear force measurements show oscillations with a fixed frequency and amplitude, see Fig.~\ref{fig:forces}. The elastic framework reproduced the primary observations---existence of two opposite moving waves, lack of a definite velocity-scale and resulting unit slip. In addition, the theory and associated numerics also provided expressions for the interface stresses, velocities and displacements. The limit $\alpha \to 0$ is exactly reproduced by the approximate functions used in the numerical scheme. The leading edge of the wave was found to resemble a stationary crack-tip with an effective remote shear stress modified by $\sV$. This correspondence, along with that between moving cracks and the onset of dynamic friction, allowed the construction of a phase diagram demarcating regions of occurrence of stick--slip and uniform sliding, see Fig.~\ref{fig:phaseDig}.

Some implications of our analytical and numerical results are now discussed. Firstly, the entire framework does not necessitate the use of an interface friction law. Indeed, the boundary value problem introduced in Sec.~\ref{sec:mechanics} applies irrespective of any friction law. Secondly, as shown in Fig.~\ref{fig:bifDig},propagation speeds for detachment waves can vary over a range of allowed values. The precise one observed depending on wave nucleation details and the width of the detachment zone. Coupled with the first implication, this means that description of slow moving waves does not need any \emph{a priori} slow velocity scale introduced into the problem via the interface friction law \cite{BrenerETAL_EurPhysJE_2005}. It must, however, be mentioned that an additional source for such a velocity could well be the material's viscoelastic response. However, incorporating this into the present framework is a formidable task and one is forced to take recourse instead to more physically motivated but algebraically simpler formulations \cite{Persson_EurPhysJE_2021}. 

Thirdly, the results for arbitrary $\nu \neq 0.5$ show that the effective \lq process zone\rq\ ahead of a moving wave is sensitive to the microscopic mechanisms at the edge of the detachment zone. This is even more important for metals and crystalline materials in general, where formation of a detachment zone via either tensile necking or compressive buckling is a difficult process. In such cases, the primary recourse to accurately determine the cut-off stress and process zone size is via molecular dynaimcs simulations \cite{WangETAL_PhysRevE_2020} and physically motivated cohesive zone models \cite{BabanETAL_PhysRevE_2020}. For compressible materials such as rubbery polymers, this is not so; infinite tensile stresses at the edges of contact are in fact common with adhesion problems in soft materials \cite{JohnsonETAL_ProcRoySocA_1971}. 

Based on our results, two very close analogies may be made for $\pm$ waves. The first is between these waves and elastic dislocations: dislocations also move under a remote shear load and cause unit (plastic) slip at the glide plane as a result. This slip is also a signed quantity in the same way that $\Delta x$ is, being always parallel to $\sV$. In fact, such an analogy, for Schallamach ($+$) waves specifically, had been speculated by Gittus in his theory of interfaceons at bimaterial interfaces \cite{Gittus_PhilMag_1975}. This work, though unaware of the occurrence of ($-$) waves at interfaces, provides a simple  model to estimate the remote stresses necessary to effect wave motion. Similar dislocation-like models have also been proposed for the failure of composite interfaces \cite{Kendall_CompInt_1996}.

The second analogy pertains to the locomotion of soft-bodied invertebrates and has already been alluded to previously \cite{ViswanathanETAL_SoftMatter_2016_1}. Since these organisms lack any limbs, they must locomote via suitable muscular movements that occur in the form of waves. Specifically, two types of waves have been identified in these organisms \cite{Trueman_SoftBodiedLocomotion_1975}. Looping locomotion is seen in caterpillars and involves a local buckle that traverses from tail to head when the organism moves forward. Likewise, retrograde waves can result from an extension of the organism's head---a tensile zone--- and traverse from head to tail. It is clear that the mechanics of these waves, effected by local muscular elasticity, have much in common with the $\pm$ waves described in this manuscript. However, putting these biological wave motions into a suitable elastic framework will involve analysis of slender objects, which is beyond the scope of the present work. It is hoped that such an analysis will also shed light on possible nucleation mechanisms applicable to the $\pm$ waves described here as well.

\section{Conclusions}

When an elastic body is slid against a rigid body at constant remote velocity, the contact interface demonstrates stick--slip motion. Our work has shown an intimate link between this intermittent interface slip and the propagation of detachment waves. Two detachment waves---Schallamach (or $+$ waves) and separation pulses (or $-$ waves) occur, with unique frequency and velocity of propagation. They move parallel ($+$) and anti-parallel ($-$) to the applied remote velocity, but cause slip in the same direction. The properties of these waves were obtained by using \emph{in situ} imaging techniques and constructing space-time diagrams.

An elastodynamic framework was presented to describe these waves theoretically. The resulting singular integral equations allowed two distinct wave solution branches, corresponding to $\pm$ waves at the interface. The interface stresses, displacements and velocities were obtained in closed form for incompressible elastic materials and a numerical scheme was used to determine these fields for more general cases. Based on these results, it was found that the leading edge of $\pm$ detachment waves resembles a stationary bimaterial interface crack. Based on this analogy, and the correspondence between interface fracture and the onset of steady sliding, a phase diagram was developed to determine when stick--slip via detachment waves would occur vis-\'a-vis steady sliding.

%\section*{Acknowledgements}

%The authors would like to acknowledge financial support from the Indian Institute of Science and the Science and Engineering Research Board (SERB), Govt. of India. 
\bibliography{bibfile}
\bibliographystyle{vancouver}
\end{document}